\documentclass{emulateapj}
\usepackage{apjfonts}
\usepackage{latexsym}
\usepackage{color}
\usepackage[usenames,dvipsnames]{xcolor}

\shorttitle{Pulsation models for SXPs}
\shortauthors{Fiorentino et al.}
\begin{document}
\def\gsim{\;\lower.6ex\hbox{$\sim$}\kern-6.7pt\raise.4ex\hbox{$>$}\;}
\def\lsim{\;\lower.6ex\hbox{$\sim$}\kern-6.7pt\raise.4ex\hbox{$<$}\;}
\title{Blue straggler masses from pulsation properties. II. Topology of the Instability Strip.}

\author{G. Fiorentino$^{1}$, M. Marconi$^{2}$, G. Bono$^{3,4}$, E. Dalessandro$^{5}$, F. R. Ferraro$^{5}$, B. Lanzoni$^{5}$, L. Lovisi$^{5}$, A. Mucciarelli$^{5}$}
\email{giuliana.fiorentino@oabo.inaf.it}
\affil{$^{1}$INAF-Osservatorio Astronomico di Bologna, via Ranzani 1, 40127, Bologna.}
\affil{$^{2}$INAF, Osservatorio Astronomico di Capodimonte,Via Moiariello 16, 80131, Napoli, Italy.}
\affil{$^{3}$Dipartimento di Fisica, Universit\`{a} di Roma Tor Vergata, Via della Ricerca Scientifica 1, 00133 Roma, Italy.}
\affil{$^{4}$INAF-Osservatorio Astronomico di Roma, Via Frascati 33, 00040, Monte Porzio Catone, Italy}
\affil{$^{5}$ Dipartimento di Fisica e Astronomia, Universit\`{a} degli Studi di Bologna, Viale Berti Pichat 6/2, 40127, Bologna, Italy}

\begin{abstract}
We present a new set of nonlinear, convective radial pulsation models
for main sequence stars computed assuming three metallicities:
Z$=$0.0001, 0.001 and 0.008. These chemical compositions bracket
the metallicity of stellar systems hosting SX Phoenicis stars (SXPs or pulsating Blue Stragglers), namely Galactic globular
clusters and nearby dwarf spheroidals. Stellar masses and luminosities
of the pulsation models are based on alpha--enhanced evolutionary
tracks from the BASTI website. We are able to define the topology of the instability strip (IS), and in turn the
pulsation relations for the first four pulsation modes. We found that
third overtones approach a stable nonlinear limit cycle. Predicted and empirical IS agree quite well
in the case of 49 SXPs belonging to $\omega$
Cen. We used theoretical Period--Luminosity
relations in B,V bands to identify their pulsation mode. We assumed Z$=$0.001 and Z$=$0.008 as mean metallicities of SXPs in $\omega$ Cen. We found respectively 13--15 fundamental, 22--6 first and 9--4 second overtone modes. Five are unstable in the third overtone mode only for Z$=$0.001. Using the above mode identification and applying the proper mass--dependent Period--Luminosity relations we found masses ranging from $\sim$1.0 to 1.2 M$_{\odot}$ ($<$M$>=$1.12, $\sigma=$0.04 M$_{\odot}$) and from $\sim$1.2 to 1.5 M$_{\odot}$ ($<$M$>=$1.33, $\sigma=$0.03 M$_{\odot}$) for Z$=$0.001
and 0.008 respectively. Our investigation supports the use of evolutionary tracks to estimate of SXP masses. We will extend our analysis to higher Helium content that may have an impact in our understanding of the BSS formation scenario.
\end{abstract}


\section{Introduction}\label{intro} 
Blue Stragglers stars (BSS) are bluer and brighter than the main--sequence (MS) turn--off (MSTO) stars. They define a sequence that in the optical Color Magnitude Diagram (CMD) can span more than 2 mag above the cluster MSTO. They mimic a younger
stellar population with masses (M=1--1.7 M$_{\odot}$, \citealt{shara97,gilliland98,demarco05,fiorentino14a}) significantly larger than normal cluster stars (MSTO$=$0.8--0.9 M$_{\odot}$). For this reason they have been suggested to be
originated from collision--induced mergers, most likely in dense stellar environments \citep{hills76,leonard89}, or by mass exchange in primordial binary systems \citep{mccrea64,zinn76,knigge09,ferraro06a,ferraro06b}. Both  formations channels can possibly be active within the same cluster \citep{ferraro09b,xin15}. SX Phoenicis (SXPs) are pulsating BSS \citep[][and reference therein]{mcnamara11} which are observed both in old-- (Globular Clusters, GCs) and intermediate (dwarf galaxies)--age stellar populations. SXPs are particularly important objects  because their pulsation properties can provide a  viable way to derive structural parameters of BSS. In particular the measure of the BSS mass is crucial since they have been recently used to define the so--called dynamical clock \citep{ferraro12} an empirical tool able to rank GCs on the basis of their dynamical ages. Since the engine of such a clock is dynamical friction and since the dynamical friction directly depends on the object mass, an accurate determination of BSS masses is of paramount importance for the calibration.\par

SXPs have luminosity amplitudes ranging from a few hundredths to tenths of magnitude, their periods 
are typically shorter than 0.1 days and can pulsate simultaneously in radial 
and in non--radial modes. They have been identified in several stellar systems, not only in Galactic GCs, but 
also in nearby dwarf irregulars (Small Magellanic Cloud, 
Large Magellanic Cloud, \citealt{soszynski02,soszynski03,poleski10}; 
IC~1613 Irr, \citealt{bernard10}) and dwarf spheroidals 
(Fornax, \citealt{poretti08}; Carina, \citealt{mateo98b,vivas13}). 
The improvement in the empirical scenario concerning SXPs was driven 
by the ongoing long term photometric surveys. The mosaic CCD cameras 
available in ground--based 4--8m class telescopes, and the Advanced Camera for 
Surveys at the Hubble Space Telescope provided the opportunity to collect 
optical, multi--band, time series data of MSTO stars of both the old 
and the intermediate--age stellar populations in the quoted stellar systems. 
The main pros of SXPs when compared with other regular variables are  
their short periods; thus SXPs can be identified even with data sets 
that only cover a few hours of observations. They are also ubiquitous,  
they have been identified in all the stellar systems 
investigated. The main con is that the luminosity amplitude is at most 
of the order of a few tenths. This means that very precise and accurate photometry down to the MSTO level is required.

\begin{figure}
\centering
\includegraphics[width=8.5cm]{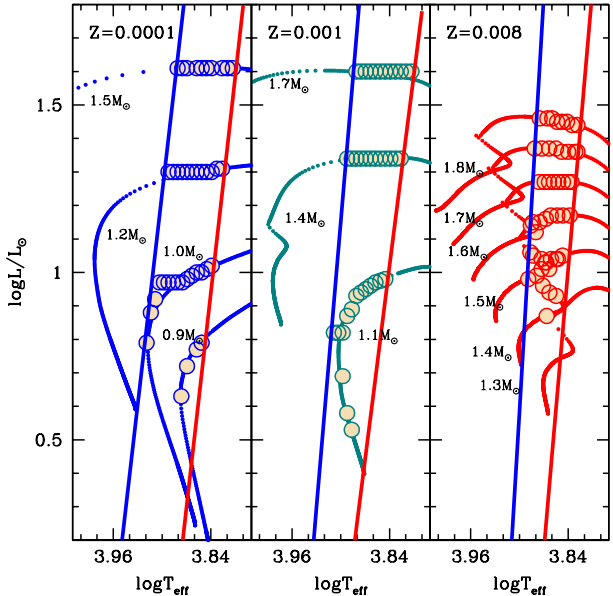}
\caption{A portion of the Hertzsprung--Russell diagram where alpha--enhanced evolutionary tracks are shown from the BASTI website \citep{pietrinferni04} for selected metallicities values that bracket values typical for Galactic GCs, e.g. Z$=$0.0001 (blue), Z$=$0.001 (teal) and Z$=$0.008 (red). For each metallicity a mass range has been selected: Z$=$0.0001, i.e. 1.0, 1.2 and 1.5  M$_{\odot}$ Z$=$0.001, i.e. 0.9, 1.1, 1.4, 1.7 M$_{\odot}$; Z$=$0.008, i.e. 1.4, 1.5, 1.6, 1.7 and 1.8 M$_{\odot}$. We also show the position in the CMD of the pulsation models (expanded dots) selected in order to roughly follow the evolutionary path. The blue and red edges derived by the pulsation instability are also shown.      
\label{fig1}}
\end{figure}

The latter empirical evidence and the fact that their oscillations are a mix 
of radial and non--radial modes are the main reasons why the pulsation 
characterization (amplitudes, modes, mean magnitudes) of cluster SXPs 
is only available for a limited sample of objects.  
However, there is solid empirical evidence \citep{mcnamara11} that non--radial modes 
become less and less relevant as soon as the stars evolve off the zero age MS. 
This means the opportunity exists to use nonlinear convective radial models to constrain 
the pulsation properties of SXPs.

This is the second paper of a series \citep[see][]{fiorentino14a} of a project 
aimed at developing a new theoretical framework for stellar oscillations in 
high gravity variables. We undertook this project, since we plan to use pulsation 
observables to constrain the intrinsic parameters of SXPs (and BSSs). In paper I \citep{fiorentino14a} we provided the pulsation 
relations to estimate the mass of SXPs using the linear, radial models provided 
by \citep{santolamazza01}. The above analysis concerning the pulsation masses of SXPs was based on the assumption 
that their pulsation mode can be constrained using the Period--Luminosity (PL) relations. 

In this paper, we further improve our analysis constructing a new set of nonlinear, 
nonlocal and time dependent convective SXP models (see Section~\ref{pulmodels}). 
This approach allows us to constrain the morphology of the instability strip (IS) 
for each assumed pulsation mode. We also describe the approach adopted to compute 
synthetic stellar populations and the pulsation relations (see Sections~\ref{pulmodels},~\ref{is} and ~\ref{relations}). 
In Section~\ref{omega} we applied the new theoretical framework to $\omega$ Centauri. 
The summary of this investigation, the further developments of the project and a few 
final remarks are given in Section~\ref{concl}.

\begin{figure}
\centering
\includegraphics[width=8.5cm]{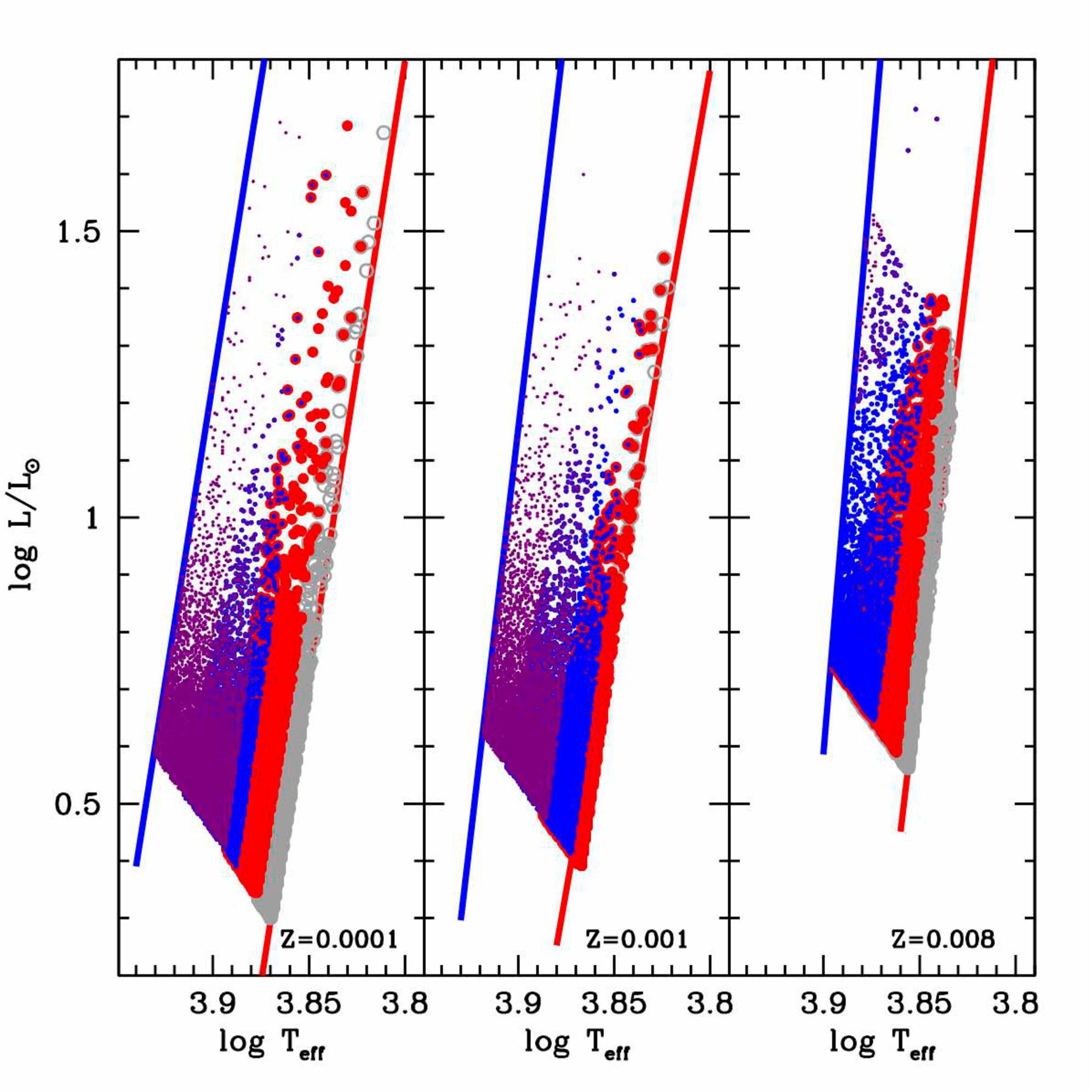}
\caption{Luminosity vs effective temperature for the synthetic populations selected according to the theoretical ISs for each pulsation mode at fixed metallicity. Fundamental, first--overtone, second--overtone and third--overtone are shown with grey, red, blue and purple symbols respectively.
\label{fig2}}
\end{figure}


\section{Evolutionary and pulsation models}\label{pulmodels} 

In this Section we present the new SXP models. To select the input stellar parameters for the nonlinear pulsation computations we considered the evolutionary tracks of the BASTI (A Bag of Stellar Tracks and Isochrones) database \citep[http://basti.oa-teramo.inaf.it/index.html, see][and references therein]{pietrinferni04,pietrinferni06} for stellar masses ranging from 1.0 to 1.8 M$_{\odot}$ and three chemical compositions, namely Z$=$0.0001 (Fe/H$=$-2.62), Z$=$0.001 (Fe/H$=$-1.62) and Z$=$0.008 (Fe/H$=$-0.70). 
The adopted evolutionary tracks are shown in Fig.~\ref{fig1} and correspond to the alpha--enhanced models \citep[see][for details]{pietrinferni06}.
For each stellar mass and chemical composition, we evaluated the pulsation instability along the corresponding evolutionary track varying the effective temperature with a step of 100 K and applying a nonlinear hydrodynamical code that includes a nonlocal time--dependent treatment of convection \citep[see][]{stellingwerf82,bono99a,bono02}. All the input parameters are given in Table~\ref{tab1}.

This code has been extensively used to investigate the properties of different classes of  evolved pulsating stars, from the low mass  RRLyrae and BL Herculis \citep[e.g.][]{bono03a,marconi03,marconi07,marconi11,dicriscienzo07} to anomalous, classical or ultra--long--period Cepheids \citep{bono00a,fiorentino02,marconi04,marconi05,fiorentino07,marconi10}. Very little has been done to interpret intermediate mass stars in the MS phase such as the metal--rich $\delta$ Scuti stars \citep{bono97c,mcnamara07} and the metal--poor SXPs \citep{gilliland98,bono02}. In this paper we investigate for the first time the nonlinear pulsation properties of SXP stars. 
The adopted physical and numerical assumptions in the hydrodynamical code are discussed in \citet{bono97a,gilliland98}. As stressed in the introduction, the importance of such nonlinear convective approach relies on the capability of predicting the complete topology of the IS and the detailed variation of all relevant stellar quantities along a pulsation cycle (e.g. light, radius and radial velocity curves). More generally, the nonlinear approach, allows us to follow the perturbed stellar envelope until the limit cycle is reached, thus also returning the amplitudes of the pulsation.

However, the computation of nonlinear radial models of SXPs is not trivial, due to their high surface gravity and very low linear growth rates \citep{bono02}. 
This means that before the models approach a pulsationally stable nonlinear limit cycle, they have to be integrated in time for a larger number of pulsation cycles, more than a factor ten time consuming than for the evolved pulsating stars (e.g. RR Lyrae, Population II Cepheids, Anomalous Cepheids).

In passing we note that the limit cycle stability for a nonlinear system would imply strictly periodic oscillations. Exact periodic solutions of the nonlinear radiative pulsation equations was introduced by \citet{stellingwerf74,stellingwerf83}. However, we still lack a similar relaxation scheme for radial oscillations taking account of a time--dependent convective transport equation.  A similar, but independent, approach was also developed by \citet{smolec08a} who decided to couple the solution of nonlinear 
conservation equations with amplitude equation formalism. This approach appears promising, but it is very time consuming. 
In this paper we use the definition that a radial mode approaches a pulsationally stable nonlinear limit cycle when the pulsation properties (period, amplitudes) over consecutive cycles attain their asymptotic behavior, in other words when the differences in the pulsation amplitudes over consecutive cycles become smaller than one part per thousand \citep{bono94b,bono99a}.This means that we are integrating the entire set of equations for a number of cycles from few to several thousands. 

This occurrence in part justifies the lack of extensive nonlinear pulsation investigations of SXPs in the literature. This has also an impact in modeling stable pulsation amplitudes and will be discussed in detail in a following paper. In this paper we will only focus on the prediction concerning the full topology of the IS for SXPs.

\subsection{Pulsation model Inputs}

In Table~\ref{tab1} we have shown the input parameters, such as mass, luminosity and effective temperature and the corresponding modes for all the modeled stars that result to be unstable for pulsation. We can use them to derive accurate pulsation relations \citep[][see Table~\ref{tab2}]{vanalbada73} and to reconstruct the full topology of the IS for the first four pulsation modes: fundamental (F), first overtone (FO), second overtone (SO) and third overtone (TO). In Fig.~\ref{fig1}, we show the {\it total} red and the blue boundaries of the IS for the fundamental and the third overtone respectively. The IS edges help us to define the portion of the Hertsprung--Russell diagram where we do expect to observe SXP stars. From an inspection of this figure, it is clear that when increasing the metallicity, the minimum mass that cross the instability strip increases. The same applies to the minimum luminosity at which we predict to observe SXPs. 

\subsection{Synthetic stellar populations}
In order to trace a more quantitative behaviour of SXPs we have created synthetic populations for each assumed metallicity using the IACstar code \citep{aparicio04}, based on the BASTI evolutionary tracks \citep{pietrinferni04}. We have assumed a constant star formation rate for $\sim$14 Gyr at each fixed chemical composition. In order to populate enough the IS we have required that at least 100,000 should be saved in the final list. This means that about 50,000,000 have been created and followed during their evolution. In this way the stars fill the whole region covered by the blue stragglers. The code returns, for each initial mass, the age, luminosity, effective temperature and gravity as well as magnitudes from U to K bands in the Johnson--Cousins photometric system \citep{bessell05}, transformed into the observational plane using colour--temperature and bolometric corrections provided by \citet{castelli04}. For each fixed metallicity, we have selected sub--populations using the four instability strips defined by our nonlinear models for each pulsation mode. To be able to compare our theoretical scenario to the observed one, we have also used transformations into the HST--WFC3 photometric system, as described in \citet{fiorentino13,fiorentino14a}. Finally, for each synthetic star we have computed the expected period by using the \citet{vanalbada73} listed in Table~\ref{tab2}.

\section{Instability strip}\label{is}
In Fig~\ref{fig2} we show the location of the synthetic stellar populations in the Hertsprung--Russel diagram, using different colours to highlight different pulsation modes. As expected, there are some regions where the synthetic stars are unstable in more than one pulsation mode.
The edges of the IS of the four pulsation modes are not parallels, thus they are not only shifted in effective temperature but they appear to define irregular regions, in particular in the case of Z$=$0.008, the highest metallicity treated in this paper. An increase in the metal content has the net effect of narrowing the {\it total} IS. Also, we note that at Z$=$0.008 the occurrence of TO pulsators tend to disappear whereas it is quite common for lower metallicities.

We have transformed the synthetic luminosities and temperatures in a number of filters from U to K \citep[][]{bessell05}, including HST WFC3 filters (F390W, F475W, F555W, F606W and F814W). 

\begin{figure*}
\centering
\includegraphics[width=8.5cm]{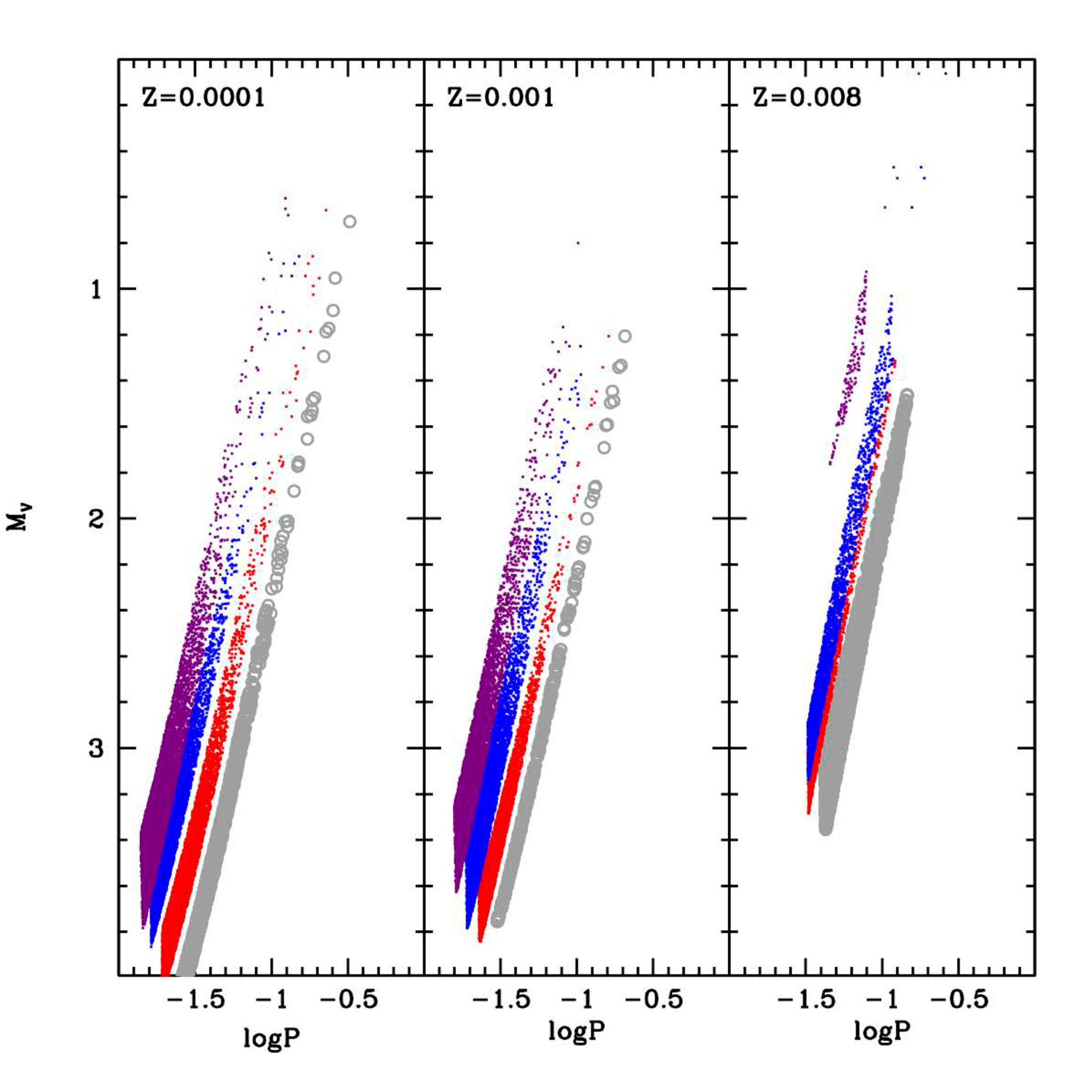}
\includegraphics[width=8.5cm]{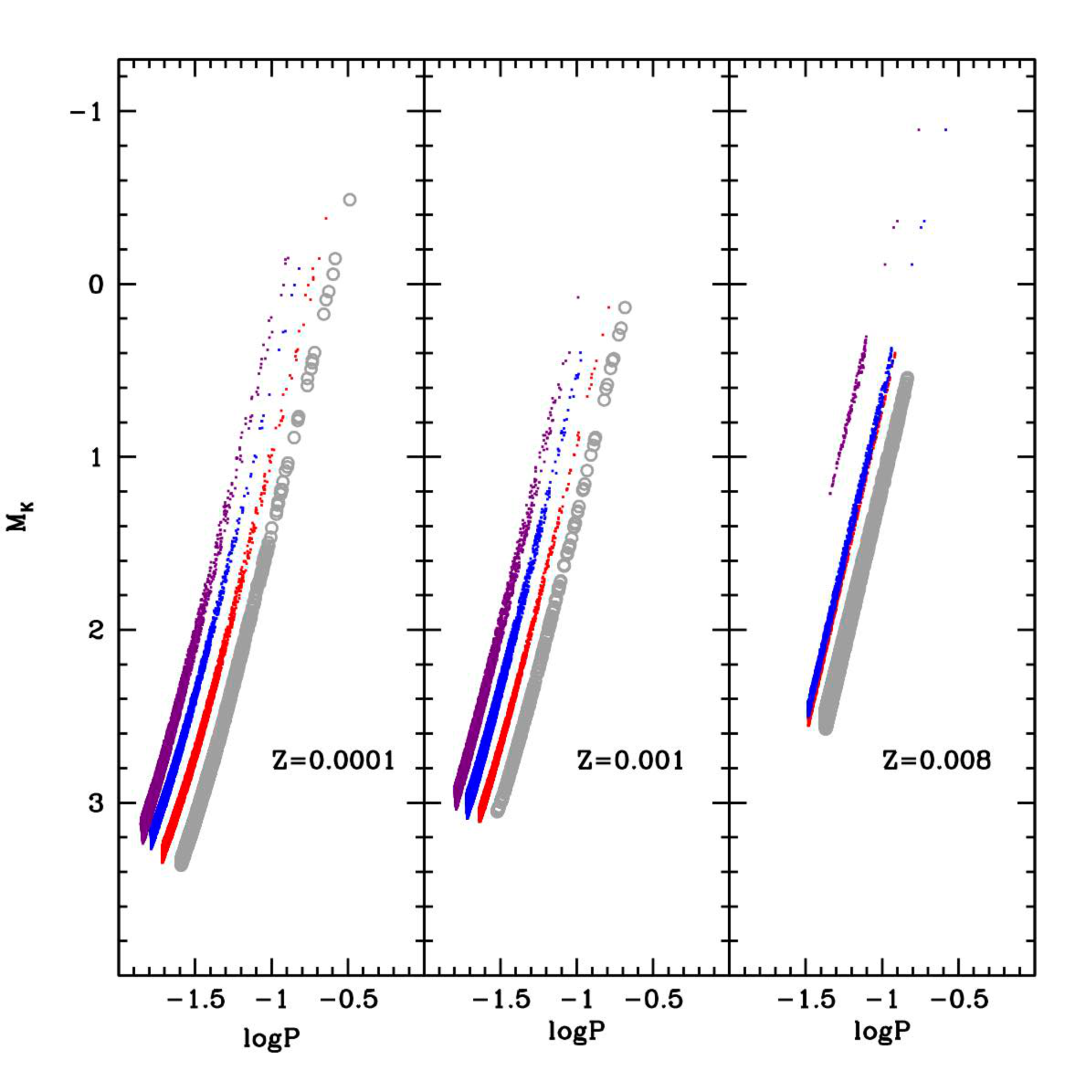}
\caption{V (left) and K--band (right) Period--Luminosity relations at varying metallicity for the first four pulsation modes. The colour code is the same as in Fig~\ref{fig2}. 
\label{fig4}}
\end{figure*}

\section{Pulsation relations}\label{relations} 
In this section, we use the synthetic populations to derive the relations between the pulsation period and the intrinsic properties of a star.  We have performed a linear fit of the magnitude as a function of the period in order to derive the so called Period--Luminosity (PL) relations. In Table~\ref{tab3} we give the PL relations for all the available photometric bands (other filters are available upon request). As an example, in Fig.~\ref{fig4} we show the PL relations in V (left) and K--band (right) for the different assumptions on the chemical compositions.
As happens for brighter and more evolved pulsators, increasing the wavelength (i.e., from V to K--band) the PL--relations become steeper and narrower (lower sigma) for each fixed metal abundance and pulsation mode. 
We notice that the slope ($\beta$) of the PL--relations become steeper when the metallicity increases, differently from what is predicted for brighter pulsators \citep[e.g., classical cepheids;][and references therein]{fiorentino13}. This behaviour is not negligible when moving to longer wavelengths. As a result we can not use a universal PL--relation to derive distances in stellar systems with very different chemical abundances. As expected, the differences in the predicted PL relations for Johnson--Cousins V, I band are negligible when compared with the corresponding filters in HST WFC3 photometric system (F555W and F814W respectively). We note that in V--band the PL relations tend to overlap making the mode separation in this plane quite difficult. Otherwise the K--band PL relations seem to be more promising in disentangling the pulsation modes. We remember here that the mode classification is a fundamental step in order to use pulsation relations for distance or mass determination \citep[see Section~\ref{omega} and discussion in ][]{fiorentino14a}.\par

The PL--relations are very easy to use, however they do suffer of some drawbacks that affect the accuracy in their use as distance indicators like the reddening correction to be applied or the intrinsic width of the IS. Partially these problems are overtaken moving to near--infrared wavelengths. However, when we use a reddening--free formulation such as the Wesenheit function they are very mitigated. In fact, these relations are not only reddening--free by definition but, including a colour term, they mimic a PL--Colour relation; thus they are narrower than the PL--relations and can be safely used for accurate distance estimates \citep[see][for a discussion]{caputo00}. Last but not least, it seems that some combinations of the filters of the Wesehneit relations are almost metallicity independent \citep{fiorentino07,inno13,braga15,marconi15}. The main con in the use of the Wesenheit relations is that their definition does depend on the adopted reddening law; for the case of RR Lyrae stars, an estimation of this effect on the distance estimate has been discussed in \citet{marconi15} and results to be less than 0.5$\%$. We have selected some optical -- near infrared  Wesenheit relations that are shown in Table~\ref{tab4}. From an inspection of this Table, it is clear that the Wesenheit relations do depend on the metallicity term, as happens for the Classical Cepheids \citep{fiorentino07}. This is different from recent finding concerning RR Lyrae stars detailed in \citet{marconi15}. In fact, the authors found a negligible dependence on the metallicity for the filter combination (B, V); however they also found that the standard deviation decreases when moving to near infrared bands, thus suggesting that a proper selection of the best relation to be used as distance indicator has to be based on our knowledge of photometric and spectroscopic uncertainties.

In the case of SXPs, we do not see any trend in the standard deviation ($\sigma_{\alpha}$) or in the metallicity coefficient ($\gamma$) of the Wesenheit relations, we can only claim that the (B, V) filter combination appears to be always the worst to be used for all the pulsation modes, whereas the remaining filter combination are quite equivalent one each other. In passing we note that the tree filter combination K, V--I is almost identical to the near infrared one K, V--K. This means that, in principle, the use of a triple band relation as distance indicator is equivalent to the dual band one.\par

We have also computed the colour--colour relations and they are shown in Table~\ref{tab5}. We highlight that the metallicity term seems not so important. When not included it would affect the standard deviation of the relations by $\sim$0.02 at the maximum.  One may notice that the intrinsic scatter of these theoretical relations is very small, with the $\sigma$(M$_B$-M$_V) \sim$ 0.01 mag, as discussed in \citet{caputo00}. The colour--colour relations can be very useful to derive an estimation of the reddening E(B--V) of the stellar systems where the SXPs are hosted, in particular by measuring the deviation of the observed colours versus the theoretical predictions.\par
Finally we provide in Table~\ref{tab6} the mass and metallicity dependent PL--relations for all the selected bands. These relations can give us fundamental constraints on the stellar mass of the BSS as extensively described in the introduction and they are the main goal of this paper. These equations can be used any time we have a detection in at least one photometric band of an SXP star, once the pulsation mode and a chemical composition are assumed. This latter case is quite obvious for most of the GCs in our Galaxy, for which accurate spectroscopic abundances are available nowadays in literature \citep[e.g.][]{carretta09}. 


\section{Theory vs observations: the case of $\omega$ Cen}\label{omega} 

\begin{figure}
\centering
\includegraphics[width=8.5cm]{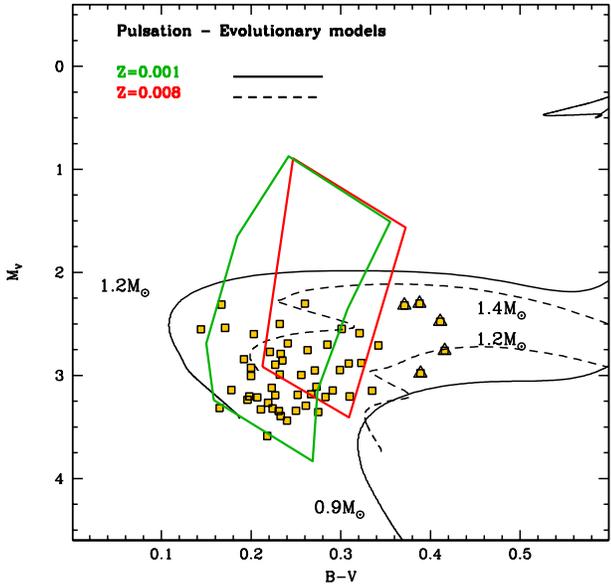}
\caption{The distribution in the CMDs of 54 SXPs (squares) in $\omega$ Cen for which we have both B and V bands from the CASE project. We also show the IS regions for Z$=$0.001 (green) and 0.008 (red). The very red (B--V $\gsim$0.36 mag) SXPs excluded from the mass--analysis are highlighted with triangles. We have also shown the  alpha enhanced evolutionary tracks for Z$=$0.001 (M=0.9 and 1.2M$_{\odot}$, solid lines) and  for Z$=$0.008 (M=1.2 and 1.4M$_{\odot}$, dashed lines) from the BASTI website \citep{pietrinferni04}.
\label{fig5}}
\end{figure}

\begin{figure*}
\centering
\includegraphics[width=8.5cm]{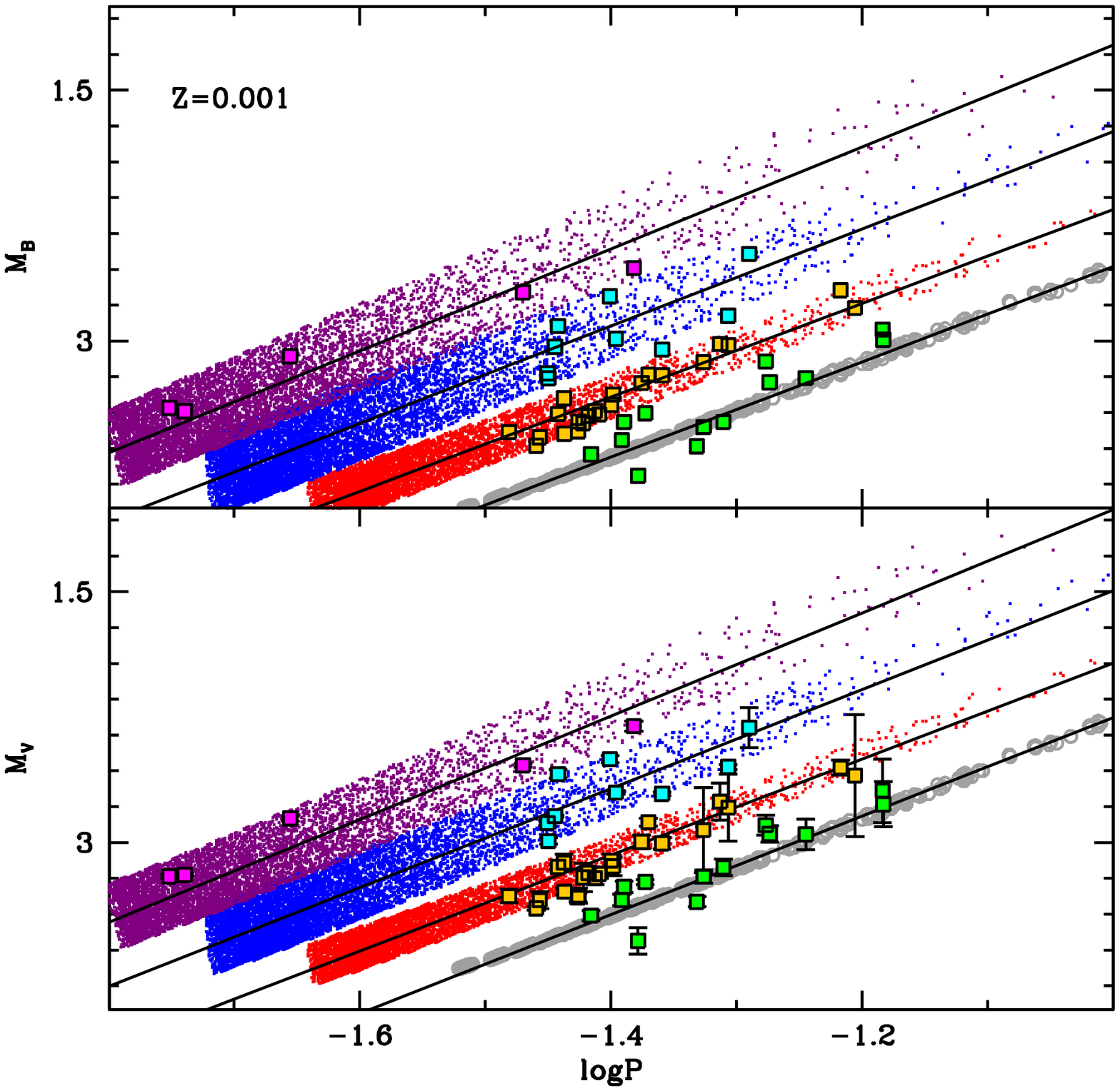}
\includegraphics[width=8.5cm]{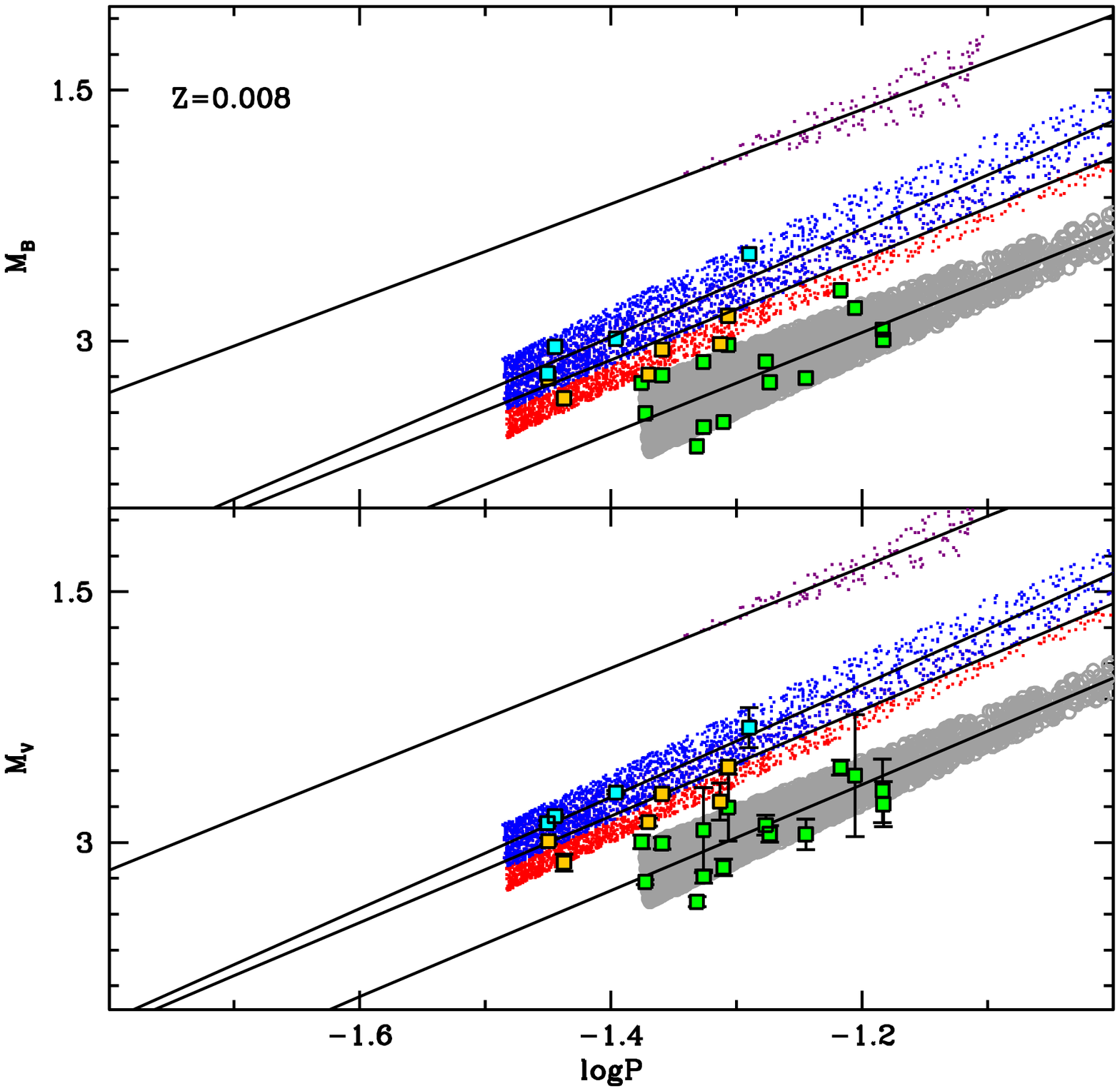}
\caption{B (top) and V--band (bottom) theoretical PL relations compared with SXPs for Z$=$0.001 (left) and 0.008 (right). Synthetic models are shown to give an idea of the PL dispersion. The pulsation modes are colour coded as in Figs~\ref{fig2} and \ref{fig4}. Magenta, cyan, orange and green indicate our classification of SXPs as third, second, first overtone and fundamental modes respectively.    
\label{fig6}}
\end{figure*}

In this section we discuss the interesting case of $\omega$ Cen, with the final aim of deriving its SXPs mass distribution. This is the most massive
and the brightest GC--like stellar system in our Galaxy, and it is characterized by a quite complex star formation history. In fact at odds with the vast majority of GC in the Galaxy, it harbors multi--populations with different iron abundance spanning a huge range of
metallicity (almost 2 dex:  -2.4 $\lsim$[Fe/H]$\lsim$ -0.4)\footnote{Note that only one other GC--like stellar system in the bulge: Terzan 5 has been
found to show such a large iron spread \citep{ferraro09a,origlia13,massari14a,massari14b}}. The properties of these sub--populations have been deeply investigated over the last 20 years using both photometry \citep{lee99,pancino00,ferraro04,sollima05,calamida09,bellini10} and spectroscopy \citep{origlia03,johnson10,pancino11b,pancino11a,monibidin12,simpson13,villanova14}. Such peculiar characteristics locate $\omega$ Cen in between the properties of GCs and the more luminous dwarf galaxies \citep{tolstoy09}. Moreover, it harbours the largest population of BSS detected so far in a stellar system (even considering only the brightest portion of the sequence, \citet{ferraro06a} identified more than 300 BSS) and possibly the largest population of SXPs. 

In particular, there are a few photometric surveys specifically dedicated to detect and characterize its variables: the two most complete ones have been carried out using CCD photometry on the 1.0--m Swope telescope at Las Campanas Observatory by the Ogle Gravitational Lensing Experiment \citep[OGLE;][]{kaluzny96} and by the Cluster Ages Experiment \citep[CASE;][]{kaluzny04,olech05}. OGLE has monitored $\omega$ Cen in V and I filters, whereas CASE in B and V filters. On the basis of their periods and their location in the CMD, the OGLE project detected and classified 24 of its variables as SXPs, whereas the CASE project increased this sample to 68 SXPs. A recent compilation of these data has been presented in \citet{cohen12} where B and V magnitudes are taken from the recent work presented in \citet{kaluzny04} and V--I colour is taken from \citet{kaluzny96}. However, the sparse sampling obtained in I band of the OGLE time--series photometry does not allow us to derive a robust V--I colour \citep[see][for a detailed discussion]{kaluzny96}; thus we decided to use only stars included in the CASE sample for which both B and V photometric bands are available; we collected a sub--sample of 54 stars.

To properly compare theory and observations we need to adopt: {\it a) } distance modulus and reddening;  {\it b) } metal content; {\it c) } pulsation modes. 

\subsection{Distance modulus}
First, we assume a distance modulus and a reddening for $\omega$ Cen, i.e. $\mu_v=$ 14.09 and E(B$-$V)$=$0.13 mag \citep[][]{kaluzny02}. This is an accurate estimation of the distance obtained using detached eclipsing double line spectroscopic binaries and it turns to be broadly consistent with recent results obtained using the RR Lyrae ($\mu_V=$ 14.18$\pm$0.04 mag) and Type II Cepheid ($\mu_V=$ 14.07$\pm$0.07 mag) stars photometered in K$s$ band from the VISTA survey \citep[][]{navarrete14,navarrete15}, that are also very similar to those of \citet{delprincipe06} and \citet{matsunaga06} based on the same distance indicators.

\subsection{Metallicity selection}
Since no direct measures of the metallicity of SXPs in Omega Centauri are available in the literature, we derived some constraints from the metallicity distribution of the BSS population. To do this we measured the metal content in 39 BSS, extracted from a larger sample including 109 BSS and studied by \citet{mucciarelli14}. The spectra were acquired by using the high-resolution spectrograph FLAMES-GIRAFFE \citep{pasquini02} at ESO-VLT, under three different programs 077.D--0696(\AA{}), 081.D--0356(\AA{}) and 089.D--0298(\AA{}), with several gratings: HR5A (R$\sim$18500, $\lambda =$ 4340--4587\AA{}), HR14A (R$\sim$18000, $\lambda =$ 6308--6701\AA{}), HR15N (R$\sim$17000, $\lambda =$ 6470--6790\AA{}), HR2 (R$\sim$19600, $\lambda =$ 3854--4049\AA{}) and HR4 (R$\sim$20350, $\lambda =$ 4340--4587\AA{}). The atmospheric parameters (effective temperature, gravity and micro-turbulence) and rotational velocities for the stars are discussed in \citet{mucciarelli14}. Due to the high rotation of many BSS in the original sample, a reliable measure of metallicity has been obtained only for 39 BSS by using 5--10 iron lines.
The metallicity distribution of these stars spans from $\sim$-2 to -0.5 with $<$[Fe/H]$>=$-1.07$\pm$0.09 ($\sigma =$ 0.6). However, the effective temperatures for most of the BSS are higher than 8000 K and thus allows the occurrence of a diffusion process called radiative levitation \citep[see ][for details]{lovisi12,lovisi13}. This process affects the surface chemical composition, mimicking a higher metal content. If we consider only
the ten coldest (T$_{eff}\lesssim$8000 K) BSS that are not greatly affected by levitation, the iron content spans from $\sim$-2 to -0.8 with $<$[Fe/H]$>=$-1.41$\pm$0.11 ($\sigma =$ 0.3). These values suggest that also SXPs may reflect the same metallicity spread of the sub-giant stellar populations.

In order to account for such a complex metallicity distribution we have decided to use the theoretical scenarios with [Fe/H]$=$-1.65 (Z$\sim$0.001)\footnote{We have verified that no significant effects on our results are caused by a metallicity change from Z$=$0.0006 to Z$=$0.001.} and  with [Fe/H]$=$-0.7 (Z$\sim$0.008). These values describe both the abundant metal--poor peak and the metal--rich tail of the $\omega$ Cen metallicity distribution \citep{johnson10}, thus bracketing the behaviour of most of the populations in this cluster, as shown in Fig.~\ref{fig5}. In this figure a direct comparison between the SXPs and the ISs for the two selected metallicities is shown. Theory and observations agree quite well with few exceptions of very red stars (B--V$\gsim$0.35 mag, five triangles) that we have excluded from the following analysis. In this figure we have over--plotted two alpha--enhanced evolutionary tracks (for each metal assumption) for masses that fully define the magnitude--colour location of the observed SXPs.

\subsection{Pulsation mode identification}

Pulsation modes are fixed using the theoretical predictions in the V--band period luminosity planes, thus they depend on the metallicity adopted, see Fig.~\ref{fig6}. The pulsation mode is in fact assigned looking at the minimum distance of each star to the four theoretical V--band PLs. In Fig.~\ref{fig6} we have also shown the B--band PL relations to check the consistency of the pulsation mode set in V--band. For the metallicity Z$=$0.001 (left panels) we can assign the following modes: 13 F, 22 FO, 9 SO and 5 TO. This is the first time that we are able to properly identify such high overtone  modes, in fact they were broadly classified as higher than F variables from \citet{mcnamara11}. Furthermore, this author derived an observational PL relation for F mode stars that turns to be much flatter ($\beta \sim$ -2.4) than the predicted ones ($\beta \sim$ -3, see Table~\ref{tab3}). This is essentially due to the different mode classification; in fact most of our F and FO are considered as F mode SXPs by McNamara (2011). However, we remember here that the distance modulus estimated applying such an observed PL relation is in very good agreement with that used in our analysis.
For Z$=0.008$ (Fig.~\ref{fig6}, right panels) we considered only those stars that fall in the region covered by models; as a consequence we have excluded many stars showing luminosities lower than predicted by pulsation and evolutionary models. In this case, we can assign a pulsation mode only for 25 stars; there are divided into 15 F, 6 FO and 4 SO.

Finally, we can use the pulsation modes for each selected metallicity to estimate masses using B and V mass--dependent PL--relations\footnote{We note here that the use of mass dependent PLC relations is not recommended and will not further discussed. This happens because the colour term turns to have a very high coefficient that changes significantly with the pulsation mode. For this reason a minimal change in the colour produces a large effect on the computed stellar mass. In particular the estimated final mass shows an unreasonable trend with pulsation mode, i.e. increasing the pulsation mode means larger masses.}. The values are given in Table~\ref{tab7} and their mean values are $<$M$> =$ 1.12 with $\sigma =$0.04 M$_{\odot}$ and  $<$M$> =$ 1.33 with $\sigma =$0.03 M$_{\odot}$ for Z$=$0.001 and Z$=$0.008 respectively, fully in agreement with the evolutionary scenario shown in Fig.~\ref{fig5} (left panel).

\section{Summary and final remarks}\label{concl} 
We developed a new and homogenous theoretical framework to account for pulsation 
properties of SXPs identified in different stellar systems, namely Galactic globulars 
and nearby dwarf galaxies. We constructed different sequences of pulsation models 
that cover a broad range in stellar masses, luminosities and in chemical compositions 
(Z=0.0001, Y=0.24; Z=0.001, Y=0.245; Z=0.008, Y=0.25). 

The pulsation scenario we developed relies on several physical assumptions. 
\begin{itemize}
\item a) The current pulsation models account for radial oscillations. However, SXPs located 
on the MS often display non-radial oscillation together with the radial ones. 

\item b) The intrinsic parameters (stellar masses and luminosities) adopted to construct 
the pulsation models are based on evolutionary tracks of single stars, while SXPs as well as the other not pulsating BSS are probably originated from the evolution of binary systems.

\item c) The pulsation relations were derived assuming a constant star formation rate in 
the computation of synthetic stellar populations (IACstar code). This means that we 
are assuming single mass evolutionary tracks to describe the behaviour of BSS. 
The last two assumptions imply that we are assuming intrinsic parameters (stellar 
masses, luminosities, chemical compositions) that are more homogeneous than observed. 
The actual properties are strongly correlated with the evolutionary status 
and the time elapsed from the formation of cluster SXps. 
\end{itemize}

The above assumptions become even more severe in dealing with SXPs in $\omega$ Cen, 
since this is one of the few globulars showing a well defined spread in iron abundance.   
However, we found that predicted and empirical properties for SXPs (49) in $\omega$ Cen 
agree quite well. The theoretical instability strips for Z$=$0.001 and 0.008 transformed into the observational 
plane (V, B--V CMD) take account of the entire sample. A few (five) red SXPs are located 
outside the predicted instability. It is not clear whether this discrepancy might be due 
to a slightly more metal-rich chemical composition or to limits in the adopted 
colour--temperature transformations or in the pulsation models close to the red edge of 
the instability strip. Note that is the region of the instability strip in which convective 
transport becomes more efficient and quenches radial oscillations.  

We provide new Period--Luminosity (PL) relations for optical and NIR bands. 
We found that the slopes of the quoted relations become steeper when 
moving from optical to NIR bands, suggesting that the latter are better diagnostics 
to estimate individual distances of SXPs. In this context it is worth mentioning that 
metal--rich SXPs, in contrast with classical Cepheids \citep{fiorentino13}, are 
brighter than metal--poor ones and their slopes are steeper, increasing the metal content.

The use of nonlinear radial models allowed us to investigate the topology of high overtones. 
We found that the occurrence of third overtones is strongly correlated with the metal 
content. They approach a stable nonlinear limit cycle in the metal-poor regime, but almost 
disappear for the most metal--rich chemical composition (Z$=$0.008). 

We used predicted PL relations in the B, V bands to perform a detailed mode identification for 
a significant fraction of SXPs in $\omega$ Cen. On the basis of the mode identification 
and of mass dependent PLZ--relations we provided an estimate of their pulsation masses. 
We found that the mean mass of SXPs does depend on the adopted chemical abundance. 
The mass distributions range from  $<$M$> =$ 1.12 ($\sigma =$0.06) M$_{\odot}$ for 
Z$=$0.001 using 49 objects to $<$M$> =$ 1.33 ($\sigma =$0.03) M$_{\odot}$ for 
Z$=$0.008 using 25 objects.  The current pulsation masses are well above the MSTO 
of canonical stellar populations in $\omega$ Cen, i.e. 0.8 M$_{\odot}$ \citep{castellani07}.

The current stellar mass distribution of SXPs in $\omega$ Cen agrees quite well 
with similar estimates based on their pulsation period \citep[see discussion in][]{mcnamara11}.
In a forthcoming investigation we plan to investigate in more detail the predicted 
pulsation properties (luminosity and velocity amplitudes) of SXPs and to compare 
them with SXPs in globulars and in nearby dwarf galaxies available in the literature. 

Moreover and even more importantly, we plan to investigate the dependence of 
the topology of the instability strip and of the pulsation properties on the 
helium content \citep{lombardi96,sills09}. Detailed hydrodynamical 
simulations of BSS formation indicate that they might have helium abundances higher 
than canonical cluster stars. 
The increase in the helium content in canonical radial pulsators causes 
a steady increase in the pulsation period, but small changes in the pulsation 
observables (amplitudes, modal stability). However, SXPs have surface gravities 
that are almost two dex larger than classical radial variables. The coupling 
between driving mechanisms, efficiency of convective transport across the 
partial ionization regions and chemical compositions needs to be investigated 
on a more quantitative basis. The preliminary comparison between theory and 
observations appears promising, a good viaticum for the forthcoming investigations.


\acknowledgments
This research is part of the project Cosmic-Lab
(http://www.cosmic-lab.eu) funded by the European Research Council
under contract ERC-2010-AdG-267675. GF has been supported also by the FIRB 2013 (grant RBFR13J716).

\bibliographystyle{apj} 

\begin{thebibliography}{92} \expandafter\ifx\csname natexlab\endcsname\relax\def\natexlab#1{#1}\fi \bibitem[{{Aparicio} \& {Gallart}(2004)}]{aparicio04} {Aparicio}, A., \& {Gallart}, C. 2004, \aj, 128, 1465 \bibitem[{{Bellini} {et~al.}(2010){Bellini}, {Bedin}, {Piotto}, {Milone},   {Marino}, \& {Villanova}}]{bellini10} {Bellini}, A., {Bedin}, L.~R., {Piotto}, G., {Milone}, A.~P., {Marino}, A.~F.,   \& {Villanova}, S. 2010, \aj, 140, 631 \bibitem[{{Bernard} {et~al.}(2010){Bernard}, {Monelli}, {Gallart}, {Aparicio},   {Cassisi}, {Drozdovsky}, {Hidalgo}, {Skillman}, \& {Stetson}}]{bernard10} {Bernard}, E.~J., {et~al.} 2010, \apj, 712, 1259 \bibitem[{{Bessell}(2005)}]{bessell05} {Bessell}, M.~S. 2005, \araa, 43, 293 \bibitem[{{Bono} {et~al.}(1997{\natexlab{a}}){Bono}, {Caputo}, {Cassisi},   {Castellani}, \& {Marconi}}]{bono97a} {Bono}, G., {Caputo}, F., {Cassisi}, S., {Castellani}, V., \& {Marconi}, M.   1997{\natexlab{a}}, \apj, 479, 279 \bibitem[{{Bono} {et~al.}(2003){Bono}, {Caputo}, {Castellani}, {Marconi},   {Storm}, \& {Degl'Innocenti}}]{bono03a} {Bono}, G., {Caputo}, F., {Castellani}, V., {Marconi}, M., {Storm}, J., \&   {Degl'Innocenti}, S. 2003, \mnras, 344, 1097 \bibitem[{{Bono} {et~al.}(2002){Bono}, {Caputo}, {Marconi}, \&   {Santolamazza}}]{bono02} {Bono}, G., {Caputo}, F., {Marconi}, M., \& {Santolamazza}, P. 2002, in   Astronomical Society of the Pacific Conference Series, Vol. 256,   Observational Aspects of Pulsating B- and A Stars, ed. C.~{Sterken} \& D.~W.   {Kurtz}, 249 \bibitem[{{Bono} {et~al.}(1997{\natexlab{b}}){Bono}, {Caputo}, {Santolamazza},   {Cassisi}, \& {Piersimoni}}]{bono97c} {Bono}, G., {Caputo}, F., {Santolamazza}, P., {Cassisi}, S., \& {Piersimoni},   A. 1997{\natexlab{b}}, \aj, 113, 2209 \bibitem[{{Bono} {et~al.}(2000){Bono}, {Castellani}, \& {Marconi}}]{bono00a} {Bono}, G., {Castellani}, V., \& {Marconi}, M. 2000, \apj, 529, 293 \bibitem[{{Bono} {et~al.}(1999){Bono}, {Marconi}, \& {Stellingwerf}}]{bono99a} {Bono}, G., {Marconi}, M., \& {Stellingwerf}, R.~F. 1999, \apjs, 122, 167 \bibitem[{{Bono} \& {Stellingwerf}(1994)}]{bono94b} {Bono}, G., \& {Stellingwerf}, R.~F. 1994, \apjs, 93, 233 \bibitem[{{Braga} {et~al.}(2015){Braga}, {Dall'Ora}, {Bono}, {Stetson},   {Ferraro}, {Iannicola}, {Marengo}, {Neeley}, {Persson}, {Buonanno},   {Coppola}, {Freedman}, {Madore}, {Marconi}, {Matsunaga}, {Monson}, {Rich},   {Scowcroft}, \& {Seibert}}]{braga15} {Braga}, V.~F., {et~al.} 2015, \apj, 799, 165 \bibitem[{{Calamida} {et~al.}(2009){Calamida}, {Bono}, {Stetson}, {Freyhammer},   {Piersimoni}, {Buonanno}, {Caputo}, {Cassisi}, {Castellani}, {Corsi},   {Dall'Ora}, {Degl'Innocenti}, {Ferraro}, {Grundahl}, {Hilker}, {Iannicola},   {Monelli}, {Nonino}, {Patat}, {Pietrinferni}, {Prada Moroni}, {Primas},   {Pulone}, {Richtler}, {Romaniello}, {Storm}, \& {Walker}}]{calamida09} {Calamida}, A., {et~al.} 2009, \apj, 706, 1277 \bibitem[{{Caputo} {et~al.}(2000){Caputo}, {Castellani}, {Marconi}, \&   {Ripepi}}]{caputo00} {Caputo}, F., {Castellani}, V., {Marconi}, M., \& {Ripepi}, V. 2000, \mnras,   316, 819 \bibitem[{{Carretta} {et~al.}(2009){Carretta}, {Bragaglia}, {Gratton},   {D'Orazi}, \& {Lucatello}}]{carretta09} {Carretta}, E., {Bragaglia}, A., {Gratton}, R., {D'Orazi}, V., \& {Lucatello},   S. 2009, \aap, 508, 695 \bibitem[{{Castellani} {et~al.}(2007){Castellani}, {Calamida}, {Bono},   {Stetson}, {Freyhammer}, {Degl'Innocenti}, {Moroni}, {Monelli}, {Corsi},   {Nonino}, {Buonanno}, {Caputo}, {Castellani}, {Dall'Ora}, {Del Principe},   {Ferraro}, {Iannicola}, {Piersimoni}, {Pulone}, \& {Vuerli}}]{castellani07} {Castellani}, V., {et~al.} 2007, \apj, 663, 1021 \bibitem[{{Castelli} \& {Kurucz}(2004)}]{castelli04} {Castelli}, F., \& {Kurucz}, R.~L. 2004, ArXiv Astrophysics e-prints \bibitem[{{Cohen} \& {Sarajedini}(2012)}]{cohen12} {Cohen}, R.~E., \& {Sarajedini}, A. 2012, \mnras, 419, 342 \bibitem[{{De Marco} {et~al.}(2005){De Marco}, {Shara}, {Zurek}, {Ouellette},   {Lanz}, {Saffer}, \& {Sepinsky}}]{demarco05} {De Marco}, O., {Shara}, M.~M., {Zurek}, D., {Ouellette}, J.~A., {Lanz}, T.,   {Saffer}, R.~A., \& {Sepinsky}, J.~F. 2005, \apj, 632, 894 \bibitem[{{Del Principe} {et~al.}(2006){Del Principe}, {Piersimoni}, {Storm},   {Caputo}, {Bono}, {Stetson}, {Castellani}, {Buonanno}, {Calamida}, {Corsi},   {Dall'Ora}, {Ferraro}, {Freyhammer}, {Iannicola}, {Monelli}, {Nonino},   {Pulone}, \& {Ripepi}}]{delprincipe06} {Del Principe}, M., {et~al.} 2006, \apj, 652, 362 \bibitem[{{Di Criscienzo} {et~al.}(2007){Di Criscienzo}, {Caputo}, {Marconi},   \& {Cassisi}}]{dicriscienzo07} {Di Criscienzo}, M., {Caputo}, F., {Marconi}, M., \& {Cassisi}, S. 2007, \aap,   471, 893 \bibitem[{{Ferraro} {et~al.}(2004){Ferraro}, {Sollima}, {Pancino},   {Bellazzini}, {Straniero}, {Origlia}, \& {Cool}}]{ferraro04} {Ferraro}, F.~R., {Sollima}, A., {Pancino}, E., {Bellazzini}, M., {Straniero},   O., {Origlia}, L., \& {Cool}, A.~M. 2004, \apjl, 603, L81 \bibitem[{{Ferraro} {et~al.}(2006{\natexlab{a}}){Ferraro}, {Sollima}, {Rood},   {Origlia}, {Pancino}, \& {Bellazzini}}]{ferraro06a} {Ferraro}, F.~R., {Sollima}, A., {Rood}, R.~T., {Origlia}, L., {Pancino}, E.,   \& {Bellazzini}, M. 2006{\natexlab{a}}, \apj, 638, 433 \bibitem[{{Ferraro} {et~al.}(2006{\natexlab{b}}){Ferraro}, {Sabbi}, {Gratton},   {Piotto}, {Lanzoni}, {Carretta}, {Rood}, {Sills}, {Fusi Pecci}, {Moehler},   {Beccari}, {Lucatello}, \& {Compagni}}]{ferraro06b} {Ferraro}, F.~R., {et~al.} 2006{\natexlab{b}}, \apjl, 647, L53 \bibitem[{{Ferraro} {et~al.}(2009{\natexlab{a}}){Ferraro}, {Dalessandro},   {Mucciarelli}, {Beccari}, {Rich}, {Origlia}, {Lanzoni}, {Rood}, {Valenti},   {Bellazzini}, {Ransom}, \& {Cocozza}}]{ferraro09a} ---. 2009{\natexlab{a}}, \nat, 462, 483 \bibitem[{{Ferraro} {et~al.}(2009{\natexlab{b}}){Ferraro}, {Beccari},   {Dalessandro}, {Lanzoni}, {Sills}, {Rood}, {Pecci}, {Karakas}, {Miocchi}, \&   {Bovinelli}}]{ferraro09b} ---. 2009{\natexlab{b}}, \nat, 462, 1028 \bibitem[{{Ferraro} {et~al.}(2012){Ferraro}, {Lanzoni}, {Dalessandro},   {Beccari}, {Pasquato}, {Miocchi}, {Rood}, {Sigurdsson}, {Sills}, {Vesperini},   {Mapelli}, {Contreras}, {Sanna}, \& {Mucciarelli}}]{ferraro12} ---. 2012, \nat, 492, 393 \bibitem[{{Fiorentino} {et~al.}(2002){Fiorentino}, {Caputo}, {Marconi}, \&   {Musella}}]{fiorentino02} {Fiorentino}, G., {Caputo}, F., {Marconi}, M., \& {Musella}, I. 2002, \apj,   576, 402 \bibitem[{{Fiorentino} {et~al.}(2014){Fiorentino}, {Lanzoni}, {Dalessandro},   {Ferraro}, {Bono}, \& {Marconi}}]{fiorentino14a} {Fiorentino}, G., {Lanzoni}, B., {Dalessandro}, E., {Ferraro}, F.~R., {Bono},   G., \& {Marconi}, M. 2014, \apj, 783, 34 \bibitem[{{Fiorentino} {et~al.}(2007){Fiorentino}, {Marconi}, {Musella}, \&   {Caputo}}]{fiorentino07} {Fiorentino}, G., {Marconi}, M., {Musella}, I., \& {Caputo}, F. 2007, \aap,   476, 863 \bibitem[{{Fiorentino} {et~al.}(2013){Fiorentino}, {Musella}, \&   {Marconi}}]{fiorentino13} {Fiorentino}, G., {Musella}, I., \& {Marconi}, M. 2013, \mnras, 434, 2866 \bibitem[{{Gilliland} {et~al.}(1998){Gilliland}, {Bono}, {Edmonds}, {Caputo},   {Cassisi}, {Petro}, {Saha}, \& {Shara}}]{gilliland98} {Gilliland}, R.~L., {Bono}, G., {Edmonds}, P.~D., {Caputo}, F., {Cassisi}, S.,   {Petro}, L.~D., {Saha}, A., \& {Shara}, M.~M. 1998, \apj, 507, 818 \bibitem[{{Hills} \& {Day}(1976)}]{hills76} {Hills}, J.~G., \& {Day}, C.~A. 1976, \aplett, 17, 87 \bibitem[{{Inno} {et~al.}(2013){Inno}, {Matsunaga}, {Bono}, {Caputo},   {Buonanno}, {Genovali}, {Laney}, {Marconi}, {Piersimoni}, {Primas}, \&   {Romaniello}}]{inno13} {Inno}, L., {et~al.} 2013, \apj, 764, 84 \bibitem[{{Johnson} \& {Pilachowski}(2010)}]{johnson10} {Johnson}, C.~I., \& {Pilachowski}, C.~A. 2010, \apj, 722, 1373 \bibitem[{{Kaluzny} {et~al.}(1996){Kaluzny}, {Kubiak}, {Szymanski}, {Udalski},   {Lrzeminski}, \& {Mateo}}]{kaluzny96} {Kaluzny}, J., {Kubiak}, M., {Szymanski}, M., {Udalski}, A., {Lrzeminski}, W.,   \& {Mateo}, M. 1996, VizieR Online Data Catalog, 412, 139 \bibitem[{{Kaluzny} {et~al.}(2004){Kaluzny}, {Olech}, {Thompson}, {Pych},   {Krzemi{\'n}ski}, \& {Schwarzenberg-Czerny}}]{kaluzny04} {Kaluzny}, J., {Olech}, A., {Thompson}, I.~B., {Pych}, W., {Krzemi{\'n}ski},   W., \& {Schwarzenberg-Czerny}, A. 2004, \aap, 424, 1101 \bibitem[{{Kaluzny} {et~al.}(2002){Kaluzny}, {Thompson}, {Krzeminski}, {Olech},   {Pych}, \& {Mochejska}}]{kaluzny02} {Kaluzny}, J., {Thompson}, I., {Krzeminski}, W., {Olech}, A., {Pych}, W., \&   {Mochejska}, B. 2002, in Astronomical Society of the Pacific Conference   Series, Vol. 265, Omega Centauri, A Unique Window into Astrophysics, ed.   F.~{van Leeuwen}, J.~D. {Hughes}, \& G.~{Piotto}, 155 \bibitem[{{Knigge} {et~al.}(2009){Knigge}, {Leigh}, \& {Sills}}]{knigge09} {Knigge}, C., {Leigh}, N., \& {Sills}, A. 2009, \nat, 457, 288 \bibitem[{{Lee} \& {Carney}(1999)}]{lee99} {Lee}, J.-W., \& {Carney}, B.~W. 1999, \aj, 118, 1373 \bibitem[{{Leonard}(1989)}]{leonard89} {Leonard}, P.~J.~T. 1989, \aj, 98, 217 \bibitem[{{Lombardi} {et~al.}(1996){Lombardi}, {Rasio}, \&   {Shapiro}}]{lombardi96} {Lombardi}, Jr., J.~C., {Rasio}, F.~A., \& {Shapiro}, S.~L. 1996, \apj, 468,   797 \bibitem[{{Lovisi} {et~al.}(2013){Lovisi}, {Mucciarelli}, {Dalessandro},   {Ferraro}, \& {Lanzoni}}]{lovisi13} {Lovisi}, L., {Mucciarelli}, A., {Dalessandro}, E., {Ferraro}, F.~R., \&   {Lanzoni}, B. 2013, \apj, 778, 64 \bibitem[{{Lovisi} {et~al.}(2012){Lovisi}, {Mucciarelli}, {Lanzoni}, {Ferraro},   {Gratton}, {Dalessandro}, \& {Contreras Ramos}}]{lovisi12} {Lovisi}, L., {Mucciarelli}, A., {Lanzoni}, B., {Ferraro}, F.~R., {Gratton},   R., {Dalessandro}, E., \& {Contreras Ramos}, R. 2012, \apj, 754, 91 \bibitem[{{Marconi} {et~al.}(2011){Marconi}, {Bono}, {Caputo}, {Piersimoni},   {Pietrinferni}, \& {Stellingwerf}}]{marconi11} {Marconi}, M., {Bono}, G., {Caputo}, F., {Piersimoni}, A.~M., {Pietrinferni},   A., \& {Stellingwerf}, R.~F. 2011, \apj, 738, 111 \bibitem[{{Marconi} {et~al.}(2003){Marconi}, {Caputo}, {Di Criscienzo}, \&   {Castellani}}]{marconi03} {Marconi}, M., {Caputo}, F., {Di Criscienzo}, M., \& {Castellani}, M. 2003,   \apj, 596, 299 \bibitem[{{Marconi} \& {Di Criscienzo}(2007)}]{marconi07} {Marconi}, M., \& {Di Criscienzo}, M. 2007, \aap, 467, 223 \bibitem[{{Marconi} {et~al.}(2004){Marconi}, {Fiorentino}, \&   {Caputo}}]{marconi04} {Marconi}, M., {Fiorentino}, G., \& {Caputo}, F. 2004, \aap, 417, 1101 \bibitem[{{Marconi} {et~al.}(2005){Marconi}, {Musella}, \&   {Fiorentino}}]{marconi05} {Marconi}, M., {Musella}, I., \& {Fiorentino}, G. 2005, \apj, 632, 590 \bibitem[{{Marconi} {et~al.}(2010){Marconi}, {Musella}, {Fiorentino},   {Clementini}, {Aloisi}, {Annibali}, {Contreras Ramos}, {Saha}, {Tosi}, \&   {van der Marel}}]{marconi10} {Marconi}, M., {et~al.} 2010, \apj, 713, 615 \bibitem[{{Marconi} {et~al.}(2015){Marconi}, {Coppola}, {Bono}, {Braga},   {Pietrinferni}, {Buonanno}, {Castellani}, {Musella}, {Ripepi}, \&   {Stellingwerf}}]{marconi15} ---. 2015, ArXiv e-prints \bibitem[{{Massari} {et~al.}(2014{\natexlab{a}}){Massari}, {Mucciarelli},   {Ferraro}, {Origlia}, {Rich}, {Lanzoni}, {Dalessandro}, {Valenti}, {Ibata},   {Lovisi}, {Bellazzini}, \& {Reitzel}}]{massari14b} {Massari}, D., {et~al.} 2014{\natexlab{a}}, \apj, 795, 22 \bibitem[{{Massari} {et~al.}(2014{\natexlab{b}}){Massari}, {Mucciarelli},   {Ferraro}, {Origlia}, {Rich}, {Lanzoni}, {Dalessandro}, {Ibata}, {Lovisi},   {Bellazzini}, \& {Reitzel}}]{massari14a} ---. 2014{\natexlab{b}}, \apj, 791, 101 \bibitem[{{Mateo} {et~al.}(1998){Mateo}, {Hurley-Keller}, \&   {Nemec}}]{mateo98b} {Mateo}, M., {Hurley-Keller}, D., \& {Nemec}, J. 1998, \aj, 115, 1856 \bibitem[{{Matsunaga} {et~al.}(2006){Matsunaga}, {Fukushi}, {Nakada},   {Tanab{\'e}}, {Feast}, {Menzies}, {Ita}, {Nishiyama}, {Baba}, {Naoi},   {Nakaya}, {Kawadu}, {Ishihara}, \& {Kato}}]{matsunaga06} {Matsunaga}, N., {et~al.} 2006, \mnras, 370, 1979 \bibitem[{{McCrea}(1964)}]{mccrea64} {McCrea}, W.~H. 1964, \mnras, 128, 147 \bibitem[{{McNamara}(2011)}]{mcnamara11} {McNamara}, D.~H. 2011, \aj, 142, 110 \bibitem[{{McNamara} {et~al.}(2007){McNamara}, {Clementini}, \&   {Marconi}}]{mcnamara07} {McNamara}, D.~H., {Clementini}, G., \& {Marconi}, M. 2007, \aj, 133, 2752 \bibitem[{{Moni Bidin} {et~al.}(2012){Moni Bidin}, {Villanova}, {Piotto},   {Moehler}, {Cassisi}, \& {Momany}}]{monibidin12} {Moni Bidin}, C., {Villanova}, S., {Piotto}, G., {Moehler}, S., {Cassisi}, S.,   \& {Momany}, Y. 2012, \aap, 547, A109 \bibitem[{{Mucciarelli} {et~al.}(2014){Mucciarelli}, {Lovisi}, {Ferraro},   {Dalessandro}, {Lanzoni}, \& {Monaco}}]{mucciarelli14} {Mucciarelli}, A., {Lovisi}, L., {Ferraro}, F.~R., {Dalessandro}, E.,   {Lanzoni}, B., \& {Monaco}, L. 2014, \apj, 797, 43 \bibitem[{{Navarrete} {et~al.}(2014){Navarrete}, {Catelan}, {Contreras Ramos},   {Gran}, {Alonso-Garc{\'{\i}}a}, \& {D{\'e}k{\'a}ny}}]{navarrete14} {Navarrete}, C., {Catelan}, M., {Contreras Ramos}, R., {Gran}, F.,   {Alonso-Garc{\'{\i}}a}, J., \& {D{\'e}k{\'a}ny}, I. 2014, in Revista Mexicana   de Astronomia y Astrofisica Conference Series, Vol.~44, Revista Mexicana de   Astronomia y Astrofisica Conference Series, 161--161 \bibitem[{{Navarrete} {et~al.}(2015){Navarrete}, {Contreras Ramos}, {Catelan},   {Clement}, {Gran}, {Alonso-Garc{\'{\i}}a}, {Angeloni}, {Hempel},   {D{\'e}k{\'a}ny}, \& {Minniti}}]{navarrete15} {Navarrete}, C., {et~al.} 2015, ArXiv e-prints \bibitem[{{Olech} {et~al.}(2005){Olech}, {Dziembowski}, {Pamyatnykh},   {Kaluzny}, {Pych}, {Schwarzenberg-Czerny}, \& {Thompson}}]{olech05} {Olech}, A., {Dziembowski}, W.~A., {Pamyatnykh}, A.~A., {Kaluzny}, J., {Pych},   W., {Schwarzenberg-Czerny}, A., \& {Thompson}, I.~B. 2005, \mnras, 363, 40 \bibitem[{{Origlia} {et~al.}(2003){Origlia}, {Ferraro}, {Bellazzini}, \&   {Pancino}}]{origlia03} {Origlia}, L., {Ferraro}, F.~R., {Bellazzini}, M., \& {Pancino}, E. 2003, \apj,   591, 916 \bibitem[{{Origlia} {et~al.}(2013){Origlia}, {Massari}, {Rich}, {Mucciarelli},   {Ferraro}, {Dalessandro}, \& {Lanzoni}}]{origlia13} {Origlia}, L., {Massari}, D., {Rich}, R.~M., {Mucciarelli}, A., {Ferraro},   F.~R., {Dalessandro}, E., \& {Lanzoni}, B. 2013, \apjl, 779, L5 \bibitem[{{Pancino} {et~al.}(2000){Pancino}, {Ferraro}, {Bellazzini}, {Piotto},   \& {Zoccali}}]{pancino00} {Pancino}, E., {Ferraro}, F.~R., {Bellazzini}, M., {Piotto}, G., \& {Zoccali},   M. 2000, \apjl, 534, L83 \bibitem[{{Pancino} {et~al.}(2011{\natexlab{a}}){Pancino}, {Mucciarelli},   {Bonifacio}, {Monaco}, \& {Sbordone}}]{pancino11b} {Pancino}, E., {Mucciarelli}, A., {Bonifacio}, P., {Monaco}, L., \& {Sbordone},   L. 2011{\natexlab{a}}, \aap, 534, A53 \bibitem[{{Pancino} {et~al.}(2011{\natexlab{b}}){Pancino}, {Mucciarelli},   {Sbordone}, {Bellazzini}, {Pasquini}, {Monaco}, \& {Ferraro}}]{pancino11a} {Pancino}, E., {Mucciarelli}, A., {Sbordone}, L., {Bellazzini}, M., {Pasquini},   L., {Monaco}, L., \& {Ferraro}, F.~R. 2011{\natexlab{b}}, \aap, 527, A18 \bibitem[{{Pasquini} {et~al.}(2002){Pasquini}, {Avila}, {Blecha}, {Cacciari},   {Cayatte}, {Colless}, {Damiani}, {de Propris}, {Dekker}, {di Marcantonio},   {Farrell}, {Gillingham}, {Guinouard}, {Hammer}, {Kaufer}, {Hill}, {Marteaud},   {Modigliani}, {Mulas}, {North}, {Popovic}, {Rossetti}, {Royer}, {Santin},   {Schmutzer}, {Simond}, {Vola}, {Waller}, \& {Zoccali}}]{pasquini02} {Pasquini}, L., {et~al.} 2002, The Messenger, 110, 1 \bibitem[{{Pietrinferni} {et~al.}(2004){Pietrinferni}, {Cassisi}, {Salaris}, \&   {Castelli}}]{pietrinferni04} {Pietrinferni}, A., {Cassisi}, S., {Salaris}, M., \& {Castelli}, F. 2004, \apj,   612, 168 \bibitem[{{Pietrinferni} {et~al.}(2006){Pietrinferni}, {Cassisi}, {Salaris}, \&   {Castelli}}]{pietrinferni06} ---. 2006, \apj, 642, 797 \bibitem[{{Poleski} {et~al.}(2010){Poleski}, {Soszy{\~n}ski}, {Udalski},   {Szyma{\~n}ski}, {Kubiak}, {Pietrzy{\~n}ski}, {Wyrzykowski}, {Szewczyk}, \&   {Ulaczyk}}]{poleski10} {Poleski}, R., {et~al.} 2010, Acta Astronomica, 60, 1 \bibitem[{{Poretti} {et~al.}(2008){Poretti}, {Clementini}, {Held}, {Greco},   {Mateo}, {Dell'Arciprete}, {Rizzi}, {Gullieuszik}, \& {Maio}}]{poretti08} {Poretti}, E., {et~al.} 2008, \apj, 685, 947 \bibitem[{{Santolamazza} {et~al.}(2001){Santolamazza}, {Marconi}, {Bono},   {Caputo}, {Cassisi}, \& {Gilliland}}]{santolamazza01} {Santolamazza}, P., {Marconi}, M., {Bono}, G., {Caputo}, F., {Cassisi}, S., \&   {Gilliland}, R.~L. 2001, \apj, 554, 1124 \bibitem[{{Shara} {et~al.}(1997){Shara}, {Saffer}, \& {Livio}}]{shara97} {Shara}, M.~M., {Saffer}, R.~A., \& {Livio}, M. 1997, \apjl, 489, L59 \bibitem[{{Sills} {et~al.}(2009){Sills}, {Karakas}, \& {Lattanzio}}]{sills09} {Sills}, A., {Karakas}, A., \& {Lattanzio}, J. 2009, \apj, 692, 1411 \bibitem[{{Simpson} \& {Cottrell}(2013)}]{simpson13} {Simpson}, J.~D., \& {Cottrell}, P.~L. 2013, \mnras, 433, 1892 \bibitem[{{Smolec} \& {Moskalik}(2008)}]{smolec08a} {Smolec}, R., \& {Moskalik}, P. 2008, Acta Astronomica, 58, 193 \bibitem[{{Sollima} {et~al.}(2005){Sollima}, {Ferraro}, {Pancino}, \&   {Bellazzini}}]{sollima05} {Sollima}, A., {Ferraro}, F.~R., {Pancino}, E., \& {Bellazzini}, M. 2005,   \mnras, 357, 265 \bibitem[{{Soszynski} {et~al.}(2002){Soszynski}, {Udalski}, {Szymanski},   {Kubiak}, {Pietrzynski}, {Wozniak}, {Zebrun}, {Szewczyk}, \&   {Wyrzykowski}}]{soszynski02} {Soszynski}, I., {et~al.} 2002, Acta Astronomica, 52, 369 \bibitem[{{Soszynski} {et~al.}(2003){Soszynski}, {Udalski}, {Szymanski},   {Kubiak}, {Pietrzynski}, {Wozniak}, {Zebrun}, {Szewczyk}, \&   {Wyrzykowski}}]{soszynski03} ---. 2003, Acta Astronomica, 53, 93 \bibitem[{{Stellingwerf}(1974)}]{stellingwerf74} {Stellingwerf}, R.~F. 1974, \apj, 192, 139 \bibitem[{{Stellingwerf}(1982)}]{stellingwerf82} ---. 1982, \apj, 262, 330 \bibitem[{{Stellingwerf}(1983)}]{stellingwerf83} ---. 1983, \apj, 271, 876 \bibitem[{{Tolstoy} {et~al.}(2009){Tolstoy}, {Hill}, \& {Tosi}}]{tolstoy09} {Tolstoy}, E., {Hill}, V., \& {Tosi}, M. 2009, \araa, 47, 371 \bibitem[{{van Albada} \& {Baker}(1973)}]{vanalbada73} {van Albada}, T.~S., \& {Baker}, N. 1973, \apj, 185, 477 \bibitem[{{Villanova} {et~al.}(2014){Villanova}, {Geisler}, {Gratton}, \&   {Cassisi}}]{villanova14} {Villanova}, S., {Geisler}, D., {Gratton}, R.~G., \& {Cassisi}, S. 2014, \apj,   791, 107 \bibitem[{{Vivas} \& {Mateo}(2013)}]{vivas13} {Vivas}, A.~K., \& {Mateo}, M. 2013, ArXiv e-prints \bibitem[{{Xin} {et~al.}(2015){Xin}, {Ferraro}, {Lu}, {Deng}, {Lanzoni},   {Dalessandro}, \& {Beccari}}]{xin15} {Xin}, Y., {Ferraro}, F.~R., {Lu}, P., {Deng}, L., {Lanzoni}, B.,   {Dalessandro}, E., \& {Beccari}, G. 2015, \apj, 801, 67 \bibitem[{{Zinn} \& {Searle}(1976)}]{zinn76} {Zinn}, R., \& {Searle}, L. 1976, \apj, 209, 734 \end{thebibliography}

\begin{center}
\begin{deluxetable*}{lclc|lclc|lclc}
\tabletypesize{\tiny}    
\tablecaption{Input parameters for the pulsation models. \label{tab1}}
\tablewidth{0pt}
\tablehead{\colhead{$\log \frac{M}{M_\odot}$}&\colhead{$T_{eff}$}&\colhead{MODE}&\colhead{$\log \frac{L}{L_{\odot}}$}&\colhead{$\log \frac{M}{M_\odot}$}&\colhead{$T_{eff}$}&\colhead{MODE}&\colhead{$\log \frac{L}{L_{\odot}}$}&\colhead{$\log \frac{M}{M_\odot}$}&\colhead{$T_{eff}$}&\colhead{MODE}&\colhead{$\log \frac{L}{L_{\odot}}$}}
\multicolumn{4}{c}{Z=0.0001, Y=0.24} & \multicolumn{4}{|c|}{Z=0.001, Y=0.245} & \multicolumn{4}{c}{Z=0.008, Y=0.25}  \\
\hline
0.9  & 7520  &  2        & 0.63     &    1.1 &  8100  &   3        &     0.82      & 1.4  &  7565  &  2   &    0.99   \\ 
0.9  & 7400  &  2        & 0.72     &    1.1 &  8000  &   3        &     0.82      & 1.4  &  7462  &  2   &    1.01   \\ 
0.9  & 7200  &  21       & 0.77     &    1.1 &  7900  &   3        &     0.82      & 1.4  &  7431  &  2   &    0.96   \\ 
0.9  & 7100  &  1        & 0.79     &    1.1 &  7800  &   3        &     0.87      & 1.4  &  7361  &  2   &    1.03   \\ 
1.0  & 8300  &  3        & 0.79     &    1.1 &  7700  &   3        &     0.89      & 1.4  &  7297  &  2   &    0.94   \\ 
1.0  & 8200  &  3        & 0.88     &    1.1 &  7600  &   23$^d$   &     0.93      & 1.4  &  7263  &  2   &    1.04   \\ 
1.0  & 8100  &  3        & 0.92     &    1.1 &  7500  &   23$^d$   &     0.94      & 1.4  &  7352  &  1   &    0.87   \\ 
1.0  & 8000  &  3        & 0.97     &    1.1 &  7400  &   23$^d$   &     0.95      & 1.4  &  7160  &  1   &    1.04   \\ 
1.0  & 7900  &  3        & 0.97     &    1.1 &  7300  &   2        &     0.96      & 1.4  &  7159  &  0   &    0.93   \\ 
1.0  & 7800  &  3        & 0.97     &    1.1 &  7200  &   2        &     0.97      & 1.4  &  7123  &  01  &    1.04   \\ 
1.0  & 7700  &  32       & 0.97     &    1.1 &  7100  &   2$^d$    &     0.975     & 1.4  &  7029  &  0   &    1.05   \\ 
1.0  & 7600  &  32       & 0.97     &    1.1 &  7000  &  01$^d$    &     0.98      & 1.5  &  7750  &   2   &   0.98     \\         
1.0  & 7500  &  32       & 0.97     &    1.1 &  7900  &  4$^{ll}$  &     0.69      & 1.5  &  7704  &   2   &   1.06     \\         
1.0  & 7400  &  1        & 0.98     &    1.1 &  7800  &  3$^{ll}$  &     0.58      & 1.5  &  7647  &   2   &   1.15     \\         
1.0  & 7300  &  1        & 0.99     &    1.1 &  7700  &  3$^{ll}$  &     0.53      & 1.5  &  7641  &   2   &   1.05     \\         
1.0  & 7200  &  1        & 1.00     &    1.4 &  7800  &  3$^{ll}$  &     1.34      & 1.5  &  7501  &   2   &   1.15     \\         
1.0  & 7100  &  1        & 1.00     &    1.4 &  7700  &  3         &     1.34      & 1.5  &  7462  &   2   &   1.04     \\         
1.0  & 7000  &  10       & 1.01     &    1.4 &  7600  &  3         &     1.34      & 1.5  &  7387  &   2   &   1.03     \\         
1.0  & 6900  &  10       & 1.02     &    1.4 &  7500  &  3         &     1.34      & 1.5  &  7372  &   2   &   1.16     \\         
1.2  & 7800  &   3       & 1.30     &    1.4 &  7400  &  3         &     1.34      & 1.5  &  7303  &  21   &   1.01     \\         
1.2  & 7700  &   3       & 1.30     &    1.4 &  7300  &  3         &     1.34      & 1.5  &  7267  &  21   &   1.17     \\         
1.2  & 7600  &   3       & 1.30     &    1.4 &  7200  &  23        &     1.34      & 1.5  &  7137  &  21   &   1.17     \\         
1.2  & 7500  &   3       & 1.30     &    1.4 &  7100  &  23$^d$    &     1.34      & 1.5  &  7020  &  10   &   1.17     \\         
1.2  & 7400  &   3       & 1.30     &    1.4 &  7000  &  12$^d$    &     1.34      & 1.5  &  6887  &  10   &   1.17     \\         
1.2  & 7300  &   3       & 1.30     &    1.4 &  6900  &  12$^d$    &     1.34      & 1.6  &  7692  &   2   &   1.14     \\         
1.2  & 7200  & 123       & 1.30     &    1.4 &  6800  &  1         &     1.34      & 1.6  &  7585  &   2   &   1.12     \\         
1.2  & 7100  &  12       & 1.30     &    1.4 &  6700  &  0$^d$     &     1.34      & 1.6  &  7513  &   3   &   1.27     \\         
1.2  & 7000  &  12       & 1.30     &    1.7 &  7600  &  3         &     1.60      & 1.6  &  7395  &  30   &   1.27     \\         
1.2  & 6900  &  12       & 1.30     &    1.7 &  7500  &  3         &     1.60      & 1.6  &  7305  &  20   &   1.27     \\         
1.2  & 6800  &   1       & 1.31     &    1.7 &  7400  &  3         &     1.60      & 1.6  &  7203  &   2$^{ll}$ &   1.27     \\    
1.2  & 6700  &  10       & 1.31     &    1.7 &  7300  &  3         &     1.60      & 1.6  &  7117  &   2$^{ll}$ &   1.27     \\    
1.5  & 7600  &   3       & 1.61     &    1.7 &  7200  &  3         &     1.60      & 1.6  &  7022  &  21   &   1.27     \\         
1.5  & 7500  &   3       & 1.61     &    1.7 &  7100  &  3         &     1.60      & 1.6  &  6922  &   1   &   1.27     \\         
1.5  & 7400  &   3       & 1.61     &    1.7 &  7000  &  3$^d$     &     1.60      & 1.6  &  6852  &   0   &   1.27     \\         
1.5  & 7200  &   3       & 1.61     &    1.7 &  6900  &  2$^d$     &     1.60      & 1.7  &  7607  &   3   &   1.37     \\         
1.5  & 7100  &  23       & 1.61     &    1.7 &  6800  &  2$^d$     &     1.60      & 1.7  &  7451  &  32   &   1.37     \\         
1.5  & 7000  &  12$^d$   & 1.61     &    1.7 &  6700  &  2$^d$     &     1.60      & 1.7  &  7302  &  32   &   1.37     \\         
1.5  & 6900  &   1       & 1.61     &    1.7 &  6600  &  0$^d$     &     1.60      & 1.7  &  7162  & 321   &   1.37     \\         
1.5  & 6700  &   0       & 1.61     &    1.7 &  6500  &  0         &     1.60      & 1.7  &  7046  & 321   &   1.36     \\         
1.5  & 6600  &   0       & 1.61     &        &        &            &               & 1.7  &  6918  &   1   &   1.36     \\         
1.5  & 6500  &   0       & 1.61     &        &        &            &               & 1.7  &  6782  &   0   &   1.36     \\         
     &       &           &          &        &        &            &               & 1.7  &  6738  &   0   &   1.36     \\         
     &       &           &          &        &        &            &               & 1.8  &  7500  &   3   &   1.46     \\         
     &       &           &          &        &        &            &               & 1.8  &  7340  &   3   &   1.46     \\         
     &       &           &          &        &        &            &               & 1.8  &  7240  &   3   &   1.46     \\         
     &       &           &          &        &        &            &               & 1.8  &  7140  &  32   &   1.45     \\         
     &       &           &          &        &        &            &               & 1.8  &  7040  &   2   &   1.45     \\         
     &       &           &          &        &        &            &               & 1.8  &  6940  &   2   &   1.45     \\         
     &       &           &          &        &        &            &               & 1.8  &  6840  &  10   &   1.44     \\         
     &       &           &          &        &        &            &               & 1.8  &  6740  &   0   &   1.44     \\         
\tablecomments{NOTE: $^d$ indicates double pulsators; $^{ll}$ indicates low--luminous pulsators, i.e. those models that for a fixed effective temperature cross the IS at a fainter luminosity level.}
\enddata
\end{deluxetable*}
\end{center}

\begin{center}
\begin{deluxetable*}{lcccccc}
\tabletypesize{\tiny}    
\tablecaption{Pulsation \citet{vanalbada73} equations for each selected pulsation modes. \label{tab2}}
\tablewidth{0pt}
\tablehead{\colhead{MODE}&\colhead{$\alpha$}&\colhead{$\beta$}&\colhead{$\gamma$}&\colhead{$\delta$}&\colhead{$\epsilon$}&\colhead{$\sigma$}}
F  &$9.331 $  &$0.7118(\pm0.0005)$  &$-2.869(\pm0.009)$  &$ -0.366(\pm0.002)$ & $ +0.0125(\pm0.0002)$ &$0.005$ \\
FO &$9.435 $  &$0.7375(\pm0.0003)$  &$-2.928(\pm0.004)$  &$ -0.425(\pm0.001)$ & $ +0.0159(\pm0.0001)$ &$0.004$ \\
SO &$10.27$   &$0.7399(\pm0.0009)$  &$-3.170(\pm0.014)$  &$ +0.043(\pm0.004)$ & $ +0.0096(\pm0.0004)$ &$0.016$ \\
TO &$10.245$  &$0.6895(\pm0.0001)$  &$-3.178(\pm0.001)$  &$ -0.292(\pm0.001)$ & $ -0.0052(\pm0.0001)$ &$0.001$
\enddata
\tablecomments{Numerical coefficients of the pulsation equations
  derived from non linear models and expressed as $\log$P$_{MODE}$ = $\alpha + \beta \times \log$L/L$_{\odot}+\gamma \log$ T$_{eff}$ $+\delta
  \log$ M/M$_{\odot}$ +$\epsilon  \log$ Z/Z$_\odot$.}
\end{deluxetable*}
\end{center}

\begin{center}
\begin{deluxetable*}{lcccccccc}
\tabletypesize{\tiny}    
\tablecaption{Period--Luminosity relations for the selected chemical abundances, pulsation modes and filters. \label{tab3}}
\tablewidth{0pt}
\tablehead{\colhead{MAG}&\colhead{$\alpha_F$}&\colhead{$\beta_F$}&\colhead{$\alpha_{FO}$}&\colhead{$\beta_{FO}$}&\colhead{$\alpha_{SO}$}&\colhead{$\beta_{SO}$}&\colhead{$\alpha_{TO}$}&\colhead{$\beta_{TO}$}}
\multicolumn{9}{c}{} \\
\multicolumn{9}{c}{Z$=$0.008} \\
\hline
\multicolumn{9}{c}{} \\
M$_U$        &-0.76$\pm$ 0.10 & -3.064$\pm$ 0.015  & -1.22$\pm$ 0.11 & -3.079$\pm$ 0.018  &  -1.61$\pm$ 0.09 & -3.273$\pm$ 0.014  &-1.91$\pm$ 0.07 & -2.930$\pm$ 0.077\\
M$_B$        &-0.68$\pm$ 0.11 & -3.025$\pm$ 0.017  & -1.13$\pm$ 0.12 & -3.031$\pm$ 0.020  &  -1.55$\pm$ 0.11 & -3.233$\pm$ 0.016  &-1.78$\pm$ 0.08 & -2.829$\pm$ 0.087\\
M$_V$        &-1.16$\pm$ 0.09 & -3.177$\pm$ 0.014  & -1.61$\pm$ 0.10 & -3.181$\pm$ 0.017  &  -1.96$\pm$ 0.09 & -3.348$\pm$ 0.013  &-2.28$\pm$ 0.07 & -3.026$\pm$ 0.072\\
M$_R$        &-1.47$\pm$ 0.08 & -3.291$\pm$ 0.012  & -1.93$\pm$ 0.08 & -3.298$\pm$ 0.014  &  -2.22$\pm$ 0.07 & -3.441$\pm$ 0.011  &-2.62$\pm$ 0.06 & -3.176$\pm$ 0.062\\
M$_I$        &-1.81$\pm$ 0.06 & -3.403$\pm$ 0.010  & -2.27$\pm$ 0.07 & -3.415$\pm$ 0.011  &  -2.51$\pm$ 0.06 & -3.535$\pm$ 0.009  &-2.97$\pm$ 0.05 & -3.326$\pm$ 0.050\\
M$_J$        &-2.17$\pm$ 0.05 & -3.541$\pm$ 0.007  & -2.63$\pm$ 0.05 & -3.553$\pm$ 0.009  &  -2.81$\pm$ 0.05 & -3.643$\pm$ 0.007  &-3.34$\pm$ 0.04 & -3.494$\pm$ 0.039\\
M$_H$        &-2.49$\pm$ 0.03 & -3.664$\pm$ 0.005  & -2.95$\pm$ 0.04 & -3.677$\pm$ 0.006  &  -3.08$\pm$ 0.03 & -3.740$\pm$ 0.005  &-3.67$\pm$ 0.03 & -3.651$\pm$ 0.027\\
M$_K$        &-2.51$\pm$ 0.03 & -3.675$\pm$ 0.005  & -2.97$\pm$ 0.04 & -3.687$\pm$ 0.006  &  -3.09$\pm$ 0.03 & -3.748$\pm$ 0.005  &-3.69$\pm$ 0.03 & -3.661$\pm$ 0.027\\
M$_{F390W}$   &-0.71$\pm$ 0.11 & -3.061$\pm$ 0.017  & -1.18$\pm$ 0.12 & -3.077$\pm$ 0.020  &  -1.61$\pm$ 0.11 & -3.293$\pm$ 0.016  &-1.86$\pm$ 0.09 & -2.883$\pm$ 0.088\\
M$_{F465W}$   &-0.97$\pm$ 0.10 & -3.148$\pm$ 0.016  & -1.44$\pm$ 0.11 & -3.155$\pm$ 0.019  &  -1.84$\pm$ 0.10 & -3.352$\pm$ 0.015  &-2.12$\pm$ 0.08 & -2.974$\pm$ 0.082\\
M$_{F555W}$   &-1.18$\pm$ 0.09 & -3.206$\pm$ 0.015  & -1.64$\pm$ 0.10 & -3.212$\pm$ 0.017  &  -2.00$\pm$ 0.09 & -3.389$\pm$ 0.014  &-2.32$\pm$ 0.07 & -3.052$\pm$ 0.075\\
M$_{F606W}$   &-1.36$\pm$ 0.08 & -3.264$\pm$ 0.013  & -1.83$\pm$ 0.09 & -3.270$\pm$ 0.016  &  -2.15$\pm$ 0.08 & -3.431$\pm$ 0.012  &-2.52$\pm$ 0.07 & -3.131$\pm$ 0.068\\
M$_{F814W}$   &-1.82$\pm$ 0.06 & -3.406$\pm$ 0.010  & -2.28$\pm$ 0.07 & -3.417$\pm$ 0.012  &  -2.53$\pm$ 0.06 & -3.539$\pm$ 0.009  &-2.99$\pm$ 0.05 & -3.329$\pm$ 0.051\\
M$_{F160W}$   &-2.45$\pm$ 0.03 & -3.631$\pm$ 0.005  & -2.91$\pm$ 0.04 & -3.644$\pm$ 0.006  &  -3.04$\pm$ 0.03 & -3.707$\pm$ 0.005  &-3.62$\pm$ 0.03 & -3.614$\pm$ 0.029\\
\hline
\multicolumn{9}{c}{} \\
\multicolumn{9}{c}{Z$=$0.001} \\
\hline
\multicolumn{9}{c}{} \\
M$_U$        &-0.39$\pm$ 0.03 & -2.861$\pm$ 0.009 & -0.55$\pm$ 0.07 & -2.731$\pm$ 0.009  &  -1.07$\pm$ 0.12 & -2.828$\pm$ 0.014   &-1.70$\pm$ 0.13 & -2.973$\pm$ 0.015 \\
M$_B$        &-0.30$\pm$ 0.02 & -2.854$\pm$ 0.005 & -0.60$\pm$ 0.08 & -2.812$\pm$ 0.011  &  -1.16$\pm$ 0.14 & -2.908$\pm$ 0.016   &-1.82$\pm$ 0.15 & -3.050$\pm$ 0.017 \\
M$_V$        &-0.70$\pm$ 0.02 & -2.952$\pm$ 0.007 & -0.95$\pm$ 0.07 & -2.875$\pm$ 0.009  &  -1.47$\pm$ 0.12 & -2.963$\pm$ 0.014   &-2.08$\pm$ 0.13 & -3.089$\pm$ 0.015 \\
M$_R$        &-0.99$\pm$ 0.02 & -3.037$\pm$ 0.007 & -1.21$\pm$ 0.06 & -2.941$\pm$ 0.008  &  -1.72$\pm$ 0.10 & -3.029$\pm$ 0.012   &-2.29$\pm$ 0.11 & -3.146$\pm$ 0.013 \\
M$_I$        &-1.31$\pm$ 0.03 & -3.127$\pm$ 0.008 & -1.50$\pm$ 0.05 & -3.013$\pm$ 0.007  &  -1.99$\pm$ 0.09 & -3.100$\pm$ 0.010   &-2.53$\pm$ 0.09 & -3.208$\pm$ 0.010 \\
M$_J$        &-1.67$\pm$ 0.03 & -3.248$\pm$ 0.008 & -1.86$\pm$ 0.04 & -3.126$\pm$ 0.005  &  -2.34$\pm$ 0.07 & -3.211$\pm$ 0.008   &-2.83$\pm$ 0.07 & -3.308$\pm$ 0.008 \\
M$_H$        &-1.99$\pm$ 0.03 & -3.351$\pm$ 0.009 & -2.14$\pm$ 0.03 & -3.207$\pm$ 0.004  &  -2.60$\pm$ 0.05 & -3.290$\pm$ 0.006   &-3.05$\pm$ 0.05 & -3.374$\pm$ 0.006 \\
M$_K$        &-2.01$\pm$ 0.03 & -3.363$\pm$ 0.009 & -2.18$\pm$ 0.03 & -3.223$\pm$ 0.004  &  -2.63$\pm$ 0.05 & -3.305$\pm$ 0.006   &-3.08$\pm$ 0.05 & -3.388$\pm$ 0.006 \\
M$_{F390W}$   &-0.33$\pm$ 0.03 & -2.877$\pm$ 0.007 & -0.56$\pm$ 0.08 & -2.786$\pm$ 0.010  &  -1.11$\pm$ 0.13 & -2.886$\pm$ 0.015   &-1.78$\pm$ 0.14 & -3.039$\pm$ 0.016 \\
M$_{F465W}$   &-0.52$\pm$ 0.03 & -2.930$\pm$ 0.007 & -0.75$\pm$ 0.08 & -2.834$\pm$ 0.010  &  -1.29$\pm$ 0.13 & -2.927$\pm$ 0.015   &-1.95$\pm$ 0.14 & -3.070$\pm$ 0.016 \\
M$_{F555W}$   &-0.70$\pm$ 0.03 & -2.970$\pm$ 0.008 & -0.92$\pm$ 0.07 & -2.867$\pm$ 0.010  &  -1.45$\pm$ 0.12 & -2.957$\pm$ 0.014   &-2.07$\pm$ 0.13 & -3.091$\pm$ 0.015 \\
M$_{F606W}$   &-0.88$\pm$ 0.03 & -3.016$\pm$ 0.008 & -1.09$\pm$ 0.07 & -2.907$\pm$ 0.009  &  -1.60$\pm$ 0.11 & -2.995$\pm$ 0.013   &-2.20$\pm$ 0.12 & -3.123$\pm$ 0.014 \\
M$_{F814W}$   &-1.33$\pm$ 0.03 & -3.134$\pm$ 0.008 & -1.51$\pm$ 0.05 & -3.013$\pm$ 0.007  &  -2.01$\pm$ 0.09 & -3.100$\pm$ 0.010   &-2.55$\pm$ 0.09 & -3.210$\pm$ 0.011 \\
M$_{F160W}$   &-1.96$\pm$ 0.03 & -3.330$\pm$ 0.009 & -2.13$\pm$ 0.03 & -3.190$\pm$ 0.004  &  -2.58$\pm$ 0.05 & -3.272$\pm$ 0.006   &-3.04$\pm$ 0.06 & -3.356$\pm$ 0.006\\
\hline
\multicolumn{9}{c}{} \\
\multicolumn{9}{c}{Z$=$0.0001} \\
\hline
\multicolumn{9}{c}{} \\
M$_U$        &-0.09$\pm$ 0.04 & -2.700$\pm$ 0.005 & -0.38$\pm$ 0.09 & -2.616$\pm$ 0.008  & -0.72$\pm$ 0.09 & -2.562$\pm$ 0.010  &-1.44$\pm$ 0.13 & -2.812$\pm$ 0.013\\
M$_B$        &-0.13$\pm$ 0.03 & -2.803$\pm$ 0.005 & -0.50$\pm$ 0.10 & -2.743$\pm$ 0.009  & -0.89$\pm$ 0.11 & -2.684$\pm$ 0.012  &-1.64$\pm$ 0.15 & -2.932$\pm$ 0.015\\
M$_V$        &-0.45$\pm$ 0.03 & -2.847$\pm$ 0.005 & -0.77$\pm$ 0.09 & -2.763$\pm$ 0.008  & -1.12$\pm$ 0.10 & -2.705$\pm$ 0.011  &-1.83$\pm$ 0.13 & -2.942$\pm$ 0.013\\
M$_R$        &-0.69$\pm$ 0.03 & -2.901$\pm$ 0.004 & -0.98$\pm$ 0.08 & -2.802$\pm$ 0.007  & -1.32$\pm$ 0.08 & -2.756$\pm$ 0.009  &-2.01$\pm$ 0.11 & -2.989$\pm$ 0.011\\
M$_I$        &-0.98$\pm$ 0.03 & -2.962$\pm$ 0.004 & -1.23$\pm$ 0.07 & -2.847$\pm$ 0.006  & -1.56$\pm$ 0.07 & -2.815$\pm$ 0.008  &-2.23$\pm$ 0.09 & -3.044$\pm$ 0.009\\
M$_J$        &-1.33$\pm$ 0.03 & -3.066$\pm$ 0.004 & -1.55$\pm$ 0.05 & -2.936$\pm$ 0.005  & -1.89$\pm$ 0.06 & -2.925$\pm$ 0.006  &-2.53$\pm$ 0.07 & -3.149$\pm$ 0.007\\
M$_H$        &-1.61$\pm$ 0.03 & -3.138$\pm$ 0.004 & -1.79$\pm$ 0.05 & -2.990$\pm$ 0.004  & -2.12$\pm$ 0.04 & -2.996$\pm$ 0.005  &-2.73$\pm$ 0.06 & -3.212$\pm$ 0.006\\
M$_K$        &-1.64$\pm$ 0.03 & -3.156$\pm$ 0.004 & -1.82$\pm$ 0.04 & -3.007$\pm$ 0.004  & -2.15$\pm$ 0.04 & -3.012$\pm$ 0.005  &-2.76$\pm$ 0.06 & -3.227$\pm$ 0.006\\
M$_{F390W}$   &-0.08$\pm$ 0.04 & -2.760$\pm$ 0.005 & -0.42$\pm$ 0.10 & -2.689$\pm$ 0.009  & -0.77$\pm$ 0.11 & -2.621$\pm$ 0.012  &-1.53$\pm$ 0.15 & -2.880$\pm$ 0.015\\
M$_{F465W}$   &-0.24$\pm$ 0.04 & -2.804$\pm$ 0.005 & -0.57$\pm$ 0.10 & -2.725$\pm$ 0.009  & -0.92$\pm$ 0.11 & -2.648$\pm$ 0.012  &-1.67$\pm$ 0.15 & -2.903$\pm$ 0.015\\
M$_{F555W}$   &-0.40$\pm$ 0.04 & -2.831$\pm$ 0.005 & -0.71$\pm$ 0.09 & -2.743$\pm$ 0.008  & -1.05$\pm$ 0.10 & -2.672$\pm$ 0.011  &-1.79$\pm$ 0.14 & -2.921$\pm$ 0.014\\
M$_{F606W}$   &-0.56$\pm$ 0.03 & -2.866$\pm$ 0.005 & -0.86$\pm$ 0.08 & -2.770$\pm$ 0.008  & -1.19$\pm$ 0.09 & -2.709$\pm$ 0.010  &-1.91$\pm$ 0.12 & -2.953$\pm$ 0.012\\
M$_{F814W}$   &-0.99$\pm$ 0.03 & -2.961$\pm$ 0.004 & -1.24$\pm$ 0.07 & -2.846$\pm$ 0.006  & -1.57$\pm$ 0.07 & -2.812$\pm$ 0.008  &-2.25$\pm$ 0.10 & -3.046$\pm$ 0.010\\
M$_{F160W}$   &-1.60$\pm$ 0.03 & -3.125$\pm$ 0.004 & -1.78$\pm$ 0.05 & -2.979$\pm$ 0.004  & -2.11$\pm$ 0.04 & -2.983$\pm$ 0.005  &-2.73$\pm$ 0.06 & -3.200$\pm$ 0.006
\enddata
\tablecomments{Numerical coefficients of the pulsation equations
  derived from non linear models and expressed as  MAG$ = \alpha + \beta \times \log$P.}
\end{deluxetable*}
\end{center}

\begin{center}
\begin{deluxetable*}{lccc}
\tabletypesize{\tiny}    
\tablecaption{metallicity dependent Wesehneit relations for each selected pulsation modes. \label{tab4}}
\tablewidth{0pt}
\tablecolumns{4}
\tablehead{\colhead{COL}&\colhead{$\alpha$}&\colhead{$\beta$}&\colhead{$\gamma$}}
\multicolumn{4}{c}{} \\
\multicolumn{4}{c}{F mode} \\
\hline
\multicolumn{4}{c}{} \\
M$_V$-3.08$\times$(B--V) & -2.51$\pm$ 0.05 & -3.267$\pm$ 0.005 & -0.1697$\pm$ 0.0009\\
M$_I$-1.43$\times$(V--I) & -2.51$\pm$ 0.05 & -3.385$\pm$ 0.005 & -0.0992$\pm$ 0.0008\\
M$_J$-0.39$\times$(V--J) & -2.37$\pm$ 0.04 & -3.380$\pm$ 0.004 & -0.0907$\pm$ 0.0007\\
M$_H$-0.21$\times$(V--H) & -2.52$\pm$ 0.04 & -3.444$\pm$ 0.004 & -0.0785$\pm$ 0.0007\\
M$_K$-0.13$\times$(V--K) & -2.47$\pm$ 0.04 & -3.430$\pm$ 0.004 & -0.0822$\pm$ 0.0007\\
M$_K$-0.27$\times$(V--I) & -2.47$\pm$ 0.04 & -3.423$\pm$ 0.004 & -0.0843$\pm$ 0.0007\\
\hline
\multicolumn{4}{c}{} \\
\multicolumn{4}{c}{FO mode} \\
\hline
\multicolumn{4}{c}{} \\
M$_V$-3.08$\times$(B--V) & -2.59$\pm$ 0.08 & -3.136$\pm$ 0.005 & -0.1345$\pm$ 0.0010\\
M$_I$-1.43$\times$(V--I) & -2.62$\pm$ 0.06 & -3.235$\pm$ 0.004 & -0.0816$\pm$ 0.0007\\
M$_J$-0.39$\times$(V--J) & -2.51$\pm$ 0.06 & -3.249$\pm$ 0.004 & -0.0711$\pm$ 0.0007\\
M$_H$-0.21$\times$(V--H) & -2.65$\pm$ 0.05 & -3.288$\pm$ 0.004 & -0.0654$\pm$ 0.0007\\
M$_K$-0.13$\times$(V--K) & -2.60$\pm$ 0.06 & -3.284$\pm$ 0.004 & -0.0667$\pm$ 0.0007\\
M$_K$-0.27$\times$(V--I) & -2.60$\pm$ 0.06 & -3.278$\pm$ 0.004 & -0.0684$\pm$ 0.0007\\
\hline
\multicolumn{4}{c}{} \\
\multicolumn{4}{c}{SO mode} \\
\hline
\multicolumn{4}{c}{} \\
M$_V$-3.08$\times$(B--V) & -2.42$\pm$ 0.07 & -3.100$\pm$ 0.004 & -0.0089$\pm$ 0.0010\\
M$_I$-1.43$\times$(V--I) & -2.46$\pm$ 0.06 & -3.173$\pm$ 0.003 &  0.0177$\pm$ 0.0008\\
M$_J$-0.39$\times$(V--J) & -2.35$\pm$ 0.06 & -3.184$\pm$ 0.004 &  0.0347$\pm$ 0.0008\\
M$_H$-0.21$\times$(V--H) & -2.48$\pm$ 0.05 & -3.211$\pm$ 0.003 &  0.0302$\pm$ 0.0008\\
M$_K$-0.13$\times$(V--K) & -2.43$\pm$ 0.06 & -3.209$\pm$ 0.003 &  0.0333$\pm$ 0.0008\\
M$_K$-0.27$\times$(V--I) & -2.42$\pm$ 0.06 & -3.204$\pm$ 0.003 &  0.0317$\pm$ 0.0008\\
\hline
\multicolumn{4}{c}{} \\
\multicolumn{4}{c}{TO mode} \\
\hline
\multicolumn{4}{c}{} \\
M$_V$-3.08$\times$(B--V) & -3.14$\pm$ 0.08 & -3.201$\pm$ 0.005 & -0.0822$\pm$ 0.0012\\
M$_I$-1.43$\times$(V--I) & -3.36$\pm$ 0.05 & -3.368$\pm$ 0.004 & -0.0674$\pm$ 0.0009\\
M$_J$-0.39$\times$(V--J) & -3.28$\pm$ 0.06 & -3.387$\pm$ 0.004 & -0.0515$\pm$ 0.0010\\
M$_H$-0.21$\times$(V--H) & -3.43$\pm$ 0.05 & -3.426$\pm$ 0.003 & -0.0578$\pm$ 0.0008\\
M$_K$-0.13$\times$(V--K) & -3.37$\pm$ 0.05 & -3.419$\pm$ 0.004 & -0.0541$\pm$ 0.0009\\
M$_K$-0.27$\times$(V--I) & -3.36$\pm$ 0.05 & -3.412$\pm$ 0.004 & -0.0554$\pm$ 0.0009
\enddata
\tablecomments{Numerical coefficients of the pulsation equations
  derived from non linear models and expressed as $\log$P = $\alpha + \beta \times \log$P$+\gamma \log$ Z.}
\end{deluxetable*}
\end{center}

\begin{center}
\begin{deluxetable*}{lccc}
\tabletypesize{\tiny}    
\tablecaption{Metallicity dependent colour--colour relations for each selected pulsation modes. \label{tab5}}
\tablewidth{0pt}
\tablecolumns{4}
\tablehead{\colhead{MODE}&\colhead{$\alpha$}&\colhead{$\beta$}&\colhead{$\gamma$}}
\multicolumn{4}{c}{} \\
\multicolumn{4}{c}{F mode} \\
\hline
\multicolumn{4}{c}{} \\
M$_V$-M$_R$ & 0.097$\pm$ 0.003 &  1.261$\pm$ 0.003 &  0.01302$\pm$ 0.00004\\
M$_V$-M$_I$ & 0.104$\pm$ 0.004 &  0.619$\pm$ 0.002 &  0.01908$\pm$ 0.00005\\
M$_V$-M$_J$ & 0.127$\pm$ 0.005 &  0.374$\pm$ 0.001 &  0.02144$\pm$ 0.00007\\
M$_V$-M$_H$ & 0.136$\pm$ 0.005 &  0.279$\pm$ 0.001 &  0.02246$\pm$ 0.00007\\
M$_V$-M$_K$ & 0.138$\pm$ 0.005 &  0.271$\pm$ 0.001 &  0.02204$\pm$ 0.00007\\
\hline
\multicolumn{4}{c}{} \\
\multicolumn{4}{c}{FO mode} \\
\hline
\multicolumn{4}{c}{} \\
M$_V$-M$_R$ & 0.110$\pm$ 0.004 &  1.134$\pm$ 0.002 &  0.01105$\pm$ 0.00004\\
M$_V$-M$_I$ & 0.118$\pm$ 0.005 &  0.541$\pm$ 0.001 &  0.01575$\pm$ 0.00005\\
M$_V$-M$_J$ & 0.135$\pm$ 0.006 &  0.329$\pm$ 0.001 &  0.01767$\pm$ 0.00006\\
M$_V$-M$_H$ & 0.144$\pm$ 0.006 &  0.244$\pm$ 0.001 &  0.01832$\pm$ 0.00006\\
M$_V$-M$_K$ & 0.145$\pm$ 0.006 &  0.239$\pm$ 0.001 &  0.01808$\pm$ 0.00006\\
\hline
\multicolumn{4}{c}{} \\
\multicolumn{4}{c}{SO mode} \\
\hline
\multicolumn{4}{c}{} \\
M$_V$-M$_R$ & 0.110$\pm$ 0.004 &  1.093$\pm$ 0.002 &  0.00978$\pm$ 0.00006\\
M$_V$-M$_I$ & 0.116$\pm$ 0.005 &  0.521$\pm$ 0.001 &  0.01398$\pm$ 0.00006\\
M$_V$-M$_J$ & 0.129$\pm$ 0.006 &  0.324$\pm$ 0.001 &  0.01610$\pm$ 0.00008\\
M$_V$-M$_H$ & 0.136$\pm$ 0.006 &  0.242$\pm$ 0.001 &  0.01656$\pm$ 0.00008\\
M$_V$-M$_K$ & 0.136$\pm$ 0.006 &  0.238$\pm$ 0.001 &  0.01648$\pm$ 0.00008\\
\hline
\multicolumn{4}{c}{} \\
\multicolumn{4}{c}{TO mode} \\
\hline
\multicolumn{4}{c}{} \\
M$_V$-M$_R$ & 0.088$\pm$ 0.005 &  1.159$\pm$ 0.002 &  0.00588$\pm$ 0.00008\\
M$_V$-M$_I$ & 0.083$\pm$ 0.005 &  0.564$\pm$ 0.001 &  0.00814$\pm$ 0.00009\\
M$_V$-M$_J$ & 0.090$\pm$ 0.007 &  0.358$\pm$ 0.001 &  0.00969$\pm$ 0.00012\\
M$_V$-M$_H$ & 0.095$\pm$ 0.007 &  0.271$\pm$ 0.001 &  0.00986$\pm$ 0.00012\\
M$_V$-M$_K$ & 0.095$\pm$ 0.007 &  0.268$\pm$ 0.001 &  0.00987$\pm$ 0.00012\\
\enddata
\tablecomments{Numerical coefficients of the pulsation equations
  derived from non linear models and expressed as B--V = $\alpha + \beta \times$COL$+\gamma \log$ Z.}
\end{deluxetable*}
\end{center}

\begin{center}
\begin{deluxetable*}{lcccc}
\tabletypesize{\tiny}    
\tablecaption{Mass--Period--Luminosity--Metallicity  relations for each selected pulsation modes. \label{tab6}}
\tablewidth{0pt}
\tablecolumns{4}
\tablehead{\colhead{MAG}&\colhead{$\alpha$}&\colhead{$\beta$}&\colhead{$\gamma$}&\colhead{$\delta$}}
\multicolumn{5}{c}{} \\
\multicolumn{5}{c}{F mode} \\
\hline
\multicolumn{5}{c}{} \\
M$_U$ &  0.19$\pm$  0.02 & -0.224$\pm$ 0.004 &  -0.62$\pm$  0.01 & 0.0628$\pm$0.0005\\
M$_B$ &  0.22$\pm$  0.03 & -0.154$\pm$ 0.004 &  -0.42$\pm$  0.01 & 0.0641$\pm$0.0007\\
M$_V$ &  0.07$\pm$  0.02 & -0.228$\pm$ 0.004 &  -0.66$\pm$  0.01 & 0.0523$\pm$0.0007\\
M$_R$ & -0.09$\pm$  0.02 & -0.295$\pm$ 0.004 &  -0.88$\pm$  0.01 & 0.0439$\pm$0.0007\\
M$_I$ & -0.31$\pm$  0.02 & -0.375$\pm$ 0.004 &  -1.16$\pm$  0.01 & 0.0380$\pm$0.0006\\
M$_J$ & -0.63$\pm$  0.02 & -0.477$\pm$ 0.004 &  -1.54$\pm$  0.01 & 0.0341$\pm$0.0005\\
M$_H$ & -0.95$\pm$  0.01 & -0.563$\pm$ 0.004 &  -1.87$\pm$  0.01 & 0.0337$\pm$0.0004\\
M$_K$ & -0.99$\pm$  0.01 & -0.573$\pm$ 0.004 &  -1.91$\pm$  0.01 & 0.0330$\pm$0.0004\\
M$_{F390W}$ &  0.21$\pm$  0.02 & -0.183$\pm$ 0.004 &  -0.50$\pm$  0.01 & 0.0628$\pm$0.0006\\
M$_{F475W}$ &  0.14$\pm$  0.02 & -0.202$\pm$ 0.004 &  -0.57$\pm$  0.01 & 0.0555$\pm$0.0007\\
M$_{F555W}$ &  0.06$\pm$  0.02 & -0.232$\pm$ 0.004 &  -0.67$\pm$  0.01 & 0.0509$\pm$0.0007\\
M$_{F606W}$ & -0.02$\pm$  0.02 & -0.266$\pm$ 0.004 &  -0.78$\pm$  0.01 & 0.0471$\pm$0.0007\\
M$_{F814W}$ & -0.31$\pm$  0.02 & -0.371$\pm$ 0.004 &  -1.14$\pm$  0.01 & 0.0387$\pm$0.0006\\
M$_{F160W}$ & -0.94$\pm$  0.01 & -0.563$\pm$ 0.004 &  -1.86$\pm$  0.01 & 0.0335$\pm$0.0004\\
\hline
\multicolumn{5}{c}{} \\
\multicolumn{5}{c}{FO mode} \\
\hline
\multicolumn{5}{c}{} \\
M$_U$ &  0.04$\pm$  0.02 & -0.238$\pm$ 0.002 &  -0.68$\pm$  0.00 & 0.0606$\pm$0.0003\\
M$_B$ &  0.06$\pm$  0.03 & -0.182$\pm$ 0.002 &  -0.54$\pm$  0.01 & 0.0610$\pm$0.0003\\
M$_V$ & -0.07$\pm$  0.02 & -0.225$\pm$ 0.002 &  -0.68$\pm$  0.01 & 0.0545$\pm$0.0003\\
M$_R$ & -0.20$\pm$  0.02 & -0.269$\pm$ 0.002 &  -0.83$\pm$  0.01 & 0.0499$\pm$0.0003\\
M$_I$ & -0.38$\pm$  0.02 & -0.329$\pm$ 0.002 &  -1.02$\pm$  0.01 & 0.0463$\pm$0.0003\\
M$_J$ & -0.66$\pm$  0.02 & -0.413$\pm$ 0.002 &  -1.32$\pm$  0.01 & 0.0432$\pm$0.0002\\
M$_H$ & -0.98$\pm$  0.01 & -0.508$\pm$ 0.001 &  -1.65$\pm$  0.00 & 0.0393$\pm$0.0002\\
M$_K$ & -1.01$\pm$  0.01 & -0.514$\pm$ 0.002 &  -1.68$\pm$  0.00 & 0.0391$\pm$0.0002\\
M$_{F390W}$ &  0.06$\pm$  0.02 & -0.201$\pm$ 0.002 &  -0.59$\pm$  0.00 & 0.0607$\pm$0.0003\\
M$_{F475W}$ &  0.00$\pm$  0.02 & -0.203$\pm$ 0.002 &  -0.61$\pm$  0.00 & 0.0565$\pm$0.0003\\
M$_{F555W}$ & -0.06$\pm$  0.02 & -0.223$\pm$ 0.002 &  -0.67$\pm$  0.00 & 0.0540$\pm$0.0003\\
M$_{F606W}$ & -0.13$\pm$  0.02 & -0.246$\pm$ 0.002 &  -0.75$\pm$  0.01 & 0.0519$\pm$0.0003\\
M$_{F814W}$ & -0.37$\pm$  0.02 & -0.324$\pm$ 0.002 &  -1.01$\pm$  0.01 & 0.0470$\pm$0.0003\\
M$_{F160W}$ & -0.94$\pm$  0.01 & -0.496$\pm$ 0.002 &  -1.61$\pm$  0.01 & 0.0410$\pm$0.0002\\
\hline
\multicolumn{5}{c}{} \\
\multicolumn{5}{c}{SO mode} \\
\hline
\multicolumn{5}{c}{} \\
M$_U$ &  0.19$\pm$  0.03 & -0.153$\pm$ 0.002 &  -0.38$\pm$  0.01 & 0.0709$\pm$0.0005\\
M$_B$ &  0.21$\pm$  0.03 & -0.120$\pm$ 0.002 &  -0.29$\pm$  0.01 & 0.0693$\pm$0.0005\\
M$_V$ &  0.13$\pm$  0.03 & -0.148$\pm$ 0.002 &  -0.38$\pm$  0.01 & 0.0682$\pm$0.0005\\
M$_R$ &  0.06$\pm$  0.03 & -0.171$\pm$ 0.002 &  -0.46$\pm$  0.01 & 0.0675$\pm$0.0005\\
M$_I$ & -0.02$\pm$  0.03 & -0.192$\pm$ 0.003 &  -0.53$\pm$  0.01 & 0.0672$\pm$0.0005\\
M$_J$ & -0.07$\pm$  0.03 & -0.190$\pm$ 0.003 &  -0.54$\pm$  0.01 & 0.0664$\pm$0.0005\\
M$_H$ & -0.05$\pm$  0.03 & -0.164$\pm$ 0.004 &  -0.46$\pm$  0.01 & 0.0641$\pm$0.0005\\
M$_K$ & -0.04$\pm$  0.03 & -0.157$\pm$ 0.004 &  -0.44$\pm$  0.01 & 0.0639$\pm$0.0005\\
M$_{F390W}$ &  0.20$\pm$  0.03 & -0.134$\pm$ 0.002 &  -0.33$\pm$  0.01 & 0.0703$\pm$0.0005\\
M$_{F475W}$ &  0.16$\pm$  0.03 & -0.140$\pm$ 0.002 &  -0.36$\pm$  0.01 & 0.0690$\pm$0.0004\\
M$_{F555W}$ &  0.13$\pm$  0.03 & -0.152$\pm$ 0.002 &  -0.40$\pm$  0.01 & 0.0685$\pm$0.0004\\
M$_{F606W}$ &  0.09$\pm$  0.03 & -0.164$\pm$ 0.002 &  -0.44$\pm$  0.01 & 0.0682$\pm$0.0004\\
M$_{F814W}$ & -0.02$\pm$  0.03 & -0.193$\pm$ 0.003 &  -0.54$\pm$  0.01 & 0.0678$\pm$0.0005\\
M$_{F160W}$ & -0.06$\pm$  0.03 & -0.171$\pm$ 0.004 &  -0.48$\pm$  0.01 & 0.0650$\pm$0.0005\\
\hline
\multicolumn{5}{c}{} \\
\multicolumn{5}{c}{TO mode} \\
\hline
\multicolumn{5}{c}{} \\
M$_U$ & -0.07$\pm$  0.02 & -0.129$\pm$ 0.001 &  -0.41$\pm$  0.00 & 0.0453$\pm$0.0004\\
M$_B$ & -0.02$\pm$  0.03 & -0.093$\pm$ 0.001 &  -0.31$\pm$  0.00 & 0.0463$\pm$0.0004\\
M$_V$ & -0.11$\pm$  0.02 & -0.121$\pm$ 0.001 &  -0.40$\pm$  0.00 & 0.0439$\pm$0.0004\\
M$_R$ & -0.21$\pm$  0.02 & -0.151$\pm$ 0.002 &  -0.50$\pm$  0.01 & 0.0415$\pm$0.0004\\
M$_I$ & -0.37$\pm$  0.02 & -0.197$\pm$ 0.002 &  -0.66$\pm$  0.01 & 0.0384$\pm$0.0004\\
M$_J$ & -0.63$\pm$  0.02 & -0.263$\pm$ 0.002 &  -0.89$\pm$  0.01 & 0.0344$\pm$0.0003\\
M$_H$ & -1.05$\pm$  0.02 & -0.375$\pm$ 0.002 &  -1.29$\pm$  0.01 & 0.0263$\pm$0.0003\\
M$_K$ & -1.06$\pm$  0.02 & -0.376$\pm$ 0.002 &  -1.30$\pm$  0.01 & 0.0263$\pm$0.0003\\
M$_{F390W}$ & -0.04$\pm$  0.02 & -0.107$\pm$ 0.001 &  -0.35$\pm$  0.00 & 0.0462$\pm$0.0004\\
M$_{F475W}$ & -0.06$\pm$  0.02 & -0.109$\pm$ 0.001 &  -0.36$\pm$  0.00 & 0.0451$\pm$0.0004\\
M$_{F555W}$ & -0.10$\pm$  0.02 & -0.121$\pm$ 0.001 &  -0.40$\pm$  0.00 & 0.0441$\pm$0.0004\\
M$_{F606W}$ & -0.16$\pm$  0.02 & -0.136$\pm$ 0.001 &  -0.45$\pm$  0.00 & 0.0430$\pm$0.0004\\
M$_{F814W}$ & -0.35$\pm$  0.02 & -0.191$\pm$ 0.002 &  -0.64$\pm$  0.01 & 0.0394$\pm$0.0004\\
M$_{F160W}$ & -0.98$\pm$  0.02 & -0.356$\pm$ 0.002 &  -1.22$\pm$  0.01 & 0.0289$\pm$0.0003\\
\enddata
\tablecomments{Numerical coefficients of the pulsation equations
  derived from non linear models and expressed as logM/M$_{\odot}$ = $\alpha + \beta \times$MAG$ + \gamma \times$log P$+\delta \log$ Z.}
\end{deluxetable*}
\end{center}

\begin{center}
\begin{deluxetable*}{cc}
\tabletypesize{\tiny}    
\tablecaption{Pulsation masses. \label{tab7}}
\tablewidth{0pt}
\tablecolumns{2}
\tablehead{\colhead{Z$=$0.001} & \colhead{Z$=$0.008}}
\multicolumn{2}{c}{}\\
\multicolumn{2}{c}{Mass (M$_{\odot}$) from B band}\\
\multicolumn{2}{c}{}\\
 $<$M(B$_{F}$)$>$~~= 1.14 $\pm$ 0.01  & $<$M(B$_{F}$)$>$~~= 1.37 $\pm$ 0.02\\ 
 $<$M(B$_{FO}$)$>$ = 1.05$\pm$ 0.01  & $<$M(B$_{FO}$)$>$ = 1.28 $\pm$ 0.02\\
 $<$M(B$_{SO}$)$>$ = 1.13 $\pm$ 0.02  & $<$M(B$_{SO}$)$>$ = 1.31 $\pm$ 0.02\\
 $<$M(B$_{TO}$)$>$ = 1.13 $\pm$ 0.02  &   \\
\hline
\multicolumn{2}{c}{}\\
\multicolumn{2}{c}{Mass (M$_{\odot}$) from V band}\\
\multicolumn{2}{c}{}\\
 $<$M(V$_{F}$)$>$  ~~= 1.14 $\pm$ 0.01   & $<$M(V$_{F}$)$>$~~= 1.36 $\pm$ 0.03 \\ 
 $<$M(V$_{FO}$)$>$ = 1.05 $\pm$ 0.01     & $<$M(V$_{FO}$)$>$ = 1.30 $\pm$ 0.02   \\
 $<$M(V$_{SO}$)$>$ = 1.16 $\pm$ 0.02     & $<$M(V$_{SO}$)$>$ = 1.34 $\pm$ 0.02   \\
 $<$M(V$_{TO}$)$>$ = 1.15 $\pm$ 0.02     & 
\enddata
\end{deluxetable*}
\end{center}


\end{document}